\documentclass[journal]{IEEEtran}
\usepackage{amsmath}
\usepackage{stmaryrd}
\usepackage{amsfonts}
\usepackage{mathrsfs}
\usepackage{amssymb,color,balance,cite}
\usepackage[dvips]{graphicx}
\usepackage{multicol}
\usepackage{mathtools}
\usepackage{caption}
\usepackage[justification=centering]{caption}
\usepackage{subfigure}
\usepackage{bbm}
\usepackage{multirow}

\newtheorem{remark}{Remark}
\newtheorem{lemma}{Lemma}
\allowdisplaybreaks[4]

\begin{document}

\title{ \Large Beamforming Network Optimization for Reducing Channel Time Variation in High-Mobility Massive MIMO
}

\author{\IEEEauthorblockN{Yinghao Ge, Weile Zhang, Feifei Gao, Shun Zhang, and Xiaoli Ma}
\thanks{This work was supported by the National Natural Science Foundation of China (NSFC) (No. 61671366 $\&$ 61601058), the Natural Science Basic Research Plan in Shaanxi Province of China (No. 2016JQ6005), and the Fundamental Research Funds for the Central Universities. The associate editor coordinating the review of this paper and approving it for publication was Cottatellucci Laura. \emph{(Corresponding author: Weile Zhang.) }

Y. Ge and W. Zhang are with the MOE Key Lab for Intelligent Networks and Network Security, School of Electronic and Information Engineering, Xi'an Jiaotong University, Xi'an, Shaanxi, 710049, China. (Emails: ge\_yinghao\_jacques@163.com, wlzhang@mail.xjtu.edu.cn).

F. Gao is with the State Key Laboratory of Intelligent Technology and Systems, Tsinghua National Laboratory for Information Science and Technology, Department of Automation, Tsinghua University, Beijing, 100084, China (Email: feifeigao@ieee.org).

S. Zhang is with the State Key Laboratory of Integrated Services Networks, Xidian University, Xi'an, Shaanxi, 710071, China (Email: zhangshunsdu@gmail.com)

X. Ma is with Georgia Institute of Technology, Atlanta, GA 30332, USA (Email: xiaoli@gatech.edu).
}
}

\maketitle

\begin{abstract}
Communications in high-mobility environments have caught a lot of attentions recently. In this paper, fast time-varying channels for massive multiple-input multiple-output (MIMO) systems are addressed. We derive the exact channel power spectrum density (PSD) for the uplink from a high-speed railway (HSR) to a base station (BS) and propose to further reduce the channel time variation via beamforming network optimization.
A large-scale uniform linear array (ULA) is equipped at the HSR to separate multiple Doppler shifts in angle domain through high-resolution transmit beamforming. Each branch comprises a dominant Doppler shift, which can be compensated to suppress the channel time variation, and we derive the channel PSD and the Doppler spread to assess the residual channel time variation.
Interestingly, the channel PSD can be exactly expressed as the product of a pattern function and a beam-distortion function. The former reflects the impact of array aperture and is the converted radiation pattern of ULA, while the latter depends on the configuration of beamforming directions.
Inspired by the PSD analysis, we introduce a common configurable amplitudes and phases (CCAP) parameter to optimize the beamforming network, by partly removing the constant modulus quantized phase constraints of matched filter (MF) beamformers. In this way, the residual Doppler shifts can be ulteriorly suppressed, further reducing the residual channel time variation. The optimal CCAP parameter minimizing the Doppler spread is derived in a closed form.
Numerical results are provided to corroborate both the channel PSD analysis and the superiority of beamforming network optimization technique.
\end{abstract}


\section{Introduction}
Over the past few decades, high-mobility communications have drawn exploding interests from researchers\cite{Wu_Access2016, He_VTM2016, Zhou_TWC2015}.
Due to the relative motion between transceivers, the emitting or incoming signals are affected by different Doppler shifts, which superimpose at the receiver and result in fast time fluctuations of the equivalent channel.
Some researchers consider the Doppler shifts as a positive factor and attempt to exploit the Doppler diversity to improve the system performance, e.g., in\cite{Wu_Access2016, Zhou_TWC2015}.
The Doppler diversity gain is harvested at the cost of high
complexity at receiver and low spectral efficiency to track the time-varying channel.
Instead, other researchers consider the fast time-varying channel detrimental to communications, since it could bring severe inter-carrier interference (ICI) to orthogonal frequency division multiplexing (OFDM) systems\cite{Hwang_TVT2009}.

When the channel is fast time-varying, it is quite challenging and even impossible to directly estimate the channel coefficients.
Some works employ the basis expansion model
(BEM)\cite{Wang_Access2018, Wang_TWC2011, Qu_TWC2010, Hijazi_TVT2009}
to approximately represent the fast time-varying channel, such that the parameters to be estimated are significantly reduced. Another frequently adopted approach approximates the channel autocorrelation as the weighted summation of two monochromatic plane waves~\cite{Souden_TSP2009, Tsai_SPL2009}.
Considering that each Doppler shift is related to an angle-of-arrival (AoA) for downlink or angle-of-departure (AoD) for uplink, the multiple Doppler shifts can be separated in angle domain. Such concept could be first found in~\cite{Norklit_TVT1999, Chizhik_TWC2004}. For reducing the channel fading rate,~\cite{Norklit_TVT1999} designs the beams that yield equal Doppler contributions by using the Fourier method, and~\cite{Chizhik_TWC2004} points out that the channel time variation can be slowed down through beamforming. These pioneering works have inspired the authors in~\cite{Zhang_ICST2011, Guo_ICSPCC2013}, where the small-scale uniform circular array (UCA) and uniform linear array (ULA) are adopted to separate multiple Doppler shifts and eliminate ICI via array beamforming. However, due to the limited spatial resolution, \cite{Zhang_ICST2011} and~\cite{Guo_ICSPCC2013} only apply to high-mobility scenarios with a few dominating paths, such as viaducts and rural areas.

In order to deal with the richly scattered high-mobility scenarios including tunnels and urban areas, we can resort to the large-scale antenna array, which is considered as a promising technique for the next generation wireless systems owing to its enhanced spectral and energy efficiency as well as high spatial resolution\cite{Rusek_SPM2013, Larsson_CM2014, Ai_JSAC2017, Zhang_TWC2018, Wang_TSP2018}.
The authors of\cite{Guo_TVT2017} propose to separate the multiple downlink Doppler shifts in angle domain by a pre-designed beamforming network with a large-scale ULA at the high-speed railway (HSR). After estimating and compensating the Doppler shift in each branch, the resultant channel turns to be quasi time-invariant and can be estimated with conventional channel estimation approaches. The array imperfection is further taken into account in \cite{Ge_VTC2017, Ge_TWC2019}, and the multi-Doppler shift separation via array beamforming can be done after array calibration.
Unlike\cite{Guo_TVT2017, Ge_VTC2017, Ge_TWC2019} addressing the downlink Doppler shifts, \cite{Guo_GLOBECOM2017} and \cite{Guo_TWC2019} focus on the uplink from the HSR to the base station (BS), where the Doppler shifts are related to AoDs instead of AoAs. As a result, a large-scale ULA is configured at the HSR to perform high-resolution transmit beamforming, and the multi-branch signal is emitted after compensating the multiple Doppler shifts in angle domain to suppress the channel time variation.
In practice, however, the number of antennas may not be sufficiently large to generate beamformers with infinite spatial resolution. Thus, the Doppler shifts cannot be completely compensated, resulting in the residual time variation of the equivalent uplink channel.
The power spectrum density (PSD) and Doppler spread are derived in\cite{Guo_GLOBECOM2017, Guo_TWC2019} as a measure of assessing such residual channel time variation, and a scaling law between the Doppler spread and number of antennas is further given in\cite{Guo_TWC2019}.
However, the PSD analysis in\cite{Guo_GLOBECOM2017, Guo_TWC2019} is approximative and only valid contingent on 1) the array is a large-scale ULA, 2) the channel follows Jakes' model\cite{Jakes_Wiley1994, Zheng_TC2003} and 3) the beamforming directions are evenly configured. The derivation is arduous and cannot be easily extended to more generalized cases.
Furthermore, the Doppler shifts separation in\cite{Guo_GLOBECOM2017, Guo_TWC2019} is performed by matched filter (MF) beamformers, which are amplitude-constrained and phase-quantized vectors\cite{Alkhateeb_JSTSP2014} and thus suboptimal in suppressing the residual Doppler shifts. Only when the number of antennas is massive, can the MF beamformers effectively eliminate the residual Doppler shifts. Otherwise, the residual channel time variation would remain non-negligible and the resultant uplink channel could still not be regarded as quasi time-invariant.

In view of this, we derive the channel PSD in an alternative way to remarkably simplify the derivation. An exact and concise expression of the channel PSD is obtained in this paper, and the derivation can be readily extended to more generalized high-mobility scenarios, where the multi-branch transmit beamforming and angle-domain Doppler shifts compensation scheme is applied. Moreover, benefiting from the simplified PSD analysis, we further propose a beamforming network optimization technique to address the suboptimality issue of MF beamformers. By introducing a common configurable amplitudes and phases (CCAP) parameter, the optimized beamformers can reduce the residual channel time variation in a more efficient manner.
The main contributions of this paper can be summarized as follows:
\begin{itemize}
  \item Explicit PSD expression with wider applicability and clearer insights: Unlike \cite{Guo_GLOBECOM2017, Guo_TWC2019} where the channel PSD is approximatively derived, we demonstrate that the channel PSD can be exactly expressed as the product of a pattern function and a beam-distortion function. The former can be uniquely determined by the antenna spacing and in fact corresponds to the radiation pattern of ULA, while the latter depends on how the beamforming directions are configured.
      Our PSD derivation not only can be extended to non-Jakes' channels or non-uniform linear arrays, but also allows to observe how the antenna spacing and beamforming directions influence the channel PSD.
  \item Reduction of Doppler spread through beamforming network optimization: The capacity of MF beamformers being limited in suppressing the residual Doppler shifts, we propose to introduce a CCAP parameter and optimize the beamforming network by removing, to some degree, the constant modulus quantized phase constraints of MF beamformers\cite{Alkhateeb_JSTSP2014}. By minimizing the corresponding Doppler spread, the optimal CCAP parameter can be acquired in a closed form. Compared to the simplest MF beamformers, the optimized beamformers can better suppress the residual channel time variation.
\end{itemize}

The rest of this paper is organized as follows. The transmit array bramforming and Doppler shifts compensation scheme under high-mobility scenarios is briefly described in Section II. Section III gives the detailed derivation of channel PSD and Doppler spread, based on which the impact of antenna spacing and beamforming directions is discussed. The beamforming network optimization technique, especially the computation of the optimal CCAP parameter, is presented in Section IV. Simulation results are provided in Section V. Section VI concludes the paper.

\textit{Notations:} Superscripts $(\cdot)^*$, $(\cdot)^T$, $(\cdot)^H$, $(\cdot)^{-1}$ and $E\{\cdot\}$ represent conjugate, transpose, Hermitian, inverse and expectation, respectively; ${\mathrm j}=\sqrt{-1}$ is the imaginary unit;
$|\cdot |$ denotes the absolute value operator;
$\|\cdot\|_{2}$ denotes the Euclidean norm of a vector or Frobenius norm of a matrix;
$\otimes$ denotes the Kronecker product operator;
$\operatorname{diag}(\mathbf x)$ is a diagonal matrix with vector $\mathbf x$ as the main diagonal;
${\mathbb C}^{m\times n}$ defines the vector space of all $m\times n$ complex matrices;
$\mathbf{I}_N$ stands for the $N\times N$ identity matrix.

\section{System Model}
\begin{figure}[t]
\setlength{\abovecaptionskip}{-1mm}
\setlength{\belowcaptionskip}{-5mm}
\begin{center}
\includegraphics[width=90mm]{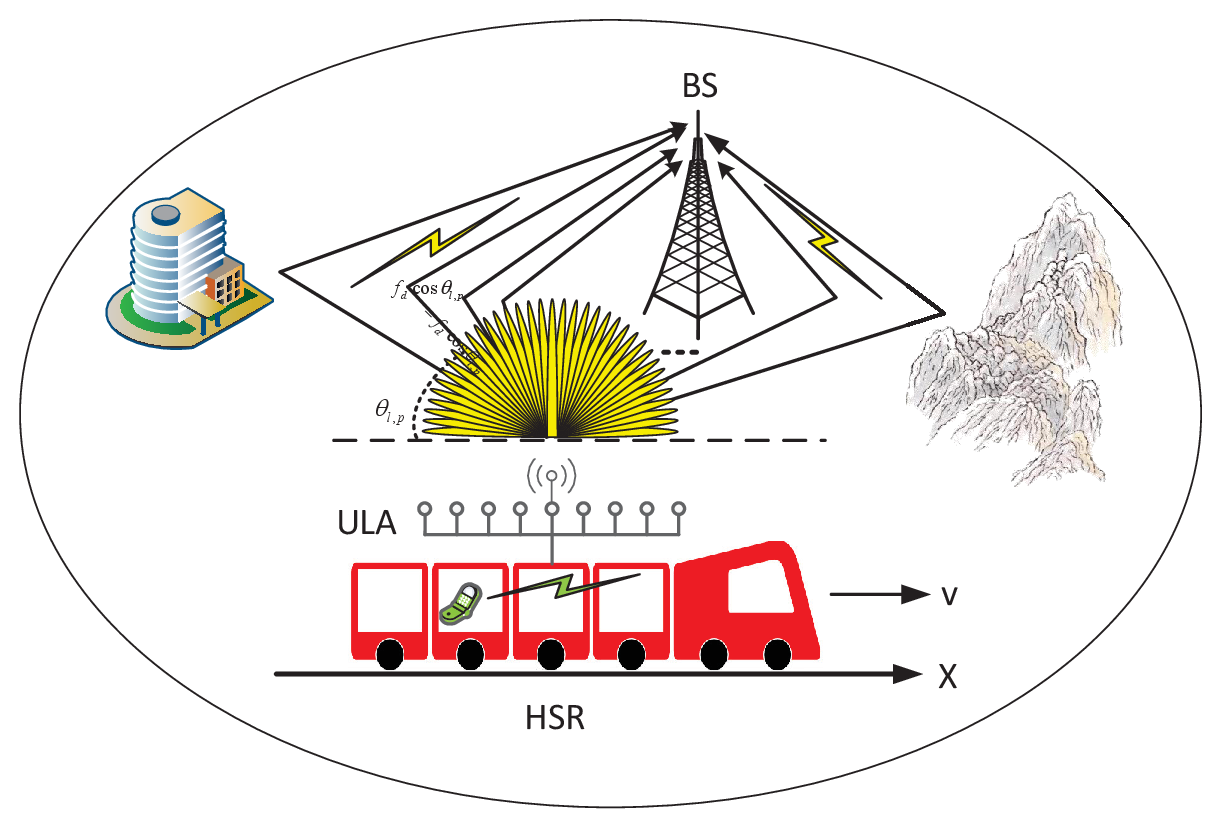}
\end{center}
\caption{ Multi-branch transmit beamforming and angle-domain Doppler shifts compensation for the high-mobility uplink. }
\end{figure}

Consider an OFDM uplink transmission in a high-mobility scenario where the signal transmitted from the HSR arrives at the BS along a number of independent subpaths, as illustrated in Fig. 1. The HSR is equipped with an $M$-elements ULA\footnote{The reason for adopting the ULA can be explained as follows (cited from~\cite{Norklit_TVT1999}): The main groups of arrays are linear, planar and circular. Compared to linear arrays, planar and circular arrays are able to generate beams to point anywhere on a sphere. Since the elevation does not influence the Doppler shift, planar or circular arrays are not needed in the considered circumstance. The ULA is the simplest array geometry and has the property of generating broader beams in the end-fire directions than at broadside, which accords with the changing rate of Doppler shifts\cite{Guo_TVT2017}.}.
Assume that the direction of the ULA coincides with that of HSR motion. Then, the array response vector pointing to direction $\theta$ can be expressed as $\mathbf{a}(\theta) \!=\! {{\Big[ \begin{matrix}
   {a_{1}(\theta)}, & {a_{2}(\theta)}, & \ldots,  & {a_{M}(\theta)}
\end{matrix}\Big]}^{T}}$, where the $r$th element is given by
${{a}_{r}}( {{\theta }}) \!=\! {{\mathrm{e}}^{\mathrm{j}2\chi (r-1) \cos {{\theta }}}}$. Here, $\chi = {\mathrm{\pi}} \frac{d}{\lambda}$, $d$ and $\lambda$ denote the antenna spacing and carrier wavelength, respectively.
By denoting the velocity of HSR as $v$, the maximum Doppler shift $f_d$ can be defined as $f_d \!=\! \frac{v}{\lambda}$. 
Note that the HSR runs along fixed tracks of the railways according to a strict preplanned schedule~\cite{Wang_TIT2015}, implying that the real-time velocity $v$ can be directly stored and accessed. Thus, the maximum Doppler shift $f_d$ is assumed perfectly known at the HSR.

The well established Jakes' channel model~\cite{Jakes_Wiley1994, Zheng_TC2003} is used to characterize the channel between the $r$th antenna and BS. It consists of $L$ taps, with $d_l$ denoting the relative delay of the $l$th tap. Each tap is composed of $P \!\gg\! 1$ separable subpaths with index $p\!=\!1,2,\ldots,P $. Denote $\theta_{l,p} \!\sim\! U\left(0, \mathrm{2\pi} \right)$ and $\rho_{l,p} \!\sim\! \mathcal{CN}\left( 0, 1/{PL} \right)$ as the departure angle and the associated complex gain of the $p$th subpath at the $l$th tap.

Denote ${{\mathbf{s}}_{m}} \!=\! {{\Big[ \begin{matrix}
   {{s}_{m}(0)}, & {{s}_{m}(1)}, & \ldots,  & {{s}_{m}(N-1)}  \\
\end{matrix} \Big]}}$ as the length-$N$ transmitted time domain symbols in the $m$th OFDM block. The cyclic prefix (CP) of length $N_{\mathrm{cp}}$ is appended to ${{\mathbf{s}}_{m}}$, which implies that $s_{m}(-n) \!=\! s_{m}(N\!-\!n)$ for $n \!=\! 1,2,\ldots, N_{\mathrm{cp}}$.
Then, the transmitted signal matrix at the transmit antenna array after delay of $d_l$ can be expressed as ${{\mathbf{S}}_{m}}\left( {{d}_{l}} \right) \!=\! {{\mathbf{1}}_{M\times 1}} \otimes {{\mathbf{s}}_{m}}\left( {{d}_{l}} \right)/ \sqrt{M}$,
where
${{\mathbf{s}}_{m}}\left( {{d}_{l}} \right) \!=\! \left[ \begin{matrix}
   {{s}_{m}}\left( -{{d}_{l}} \right), & {{s}_{m}}\left( 1 \!-\! {{d}_{l}} \right), & \ldots, & {{s}_{m}}\left( N\!-\!1\!-\!{{d}_{l}} \right)  \\
\end{matrix} \right]$ corresponds to the right circular shift of ${{\mathbf{s}}_{m}}$ by a factor of $d_l$.
Here, the divisor $\sqrt{M}$ is added to keep the total transmit power per symbol to 1.
Moreover, define $N_s \!=\! N \!+\! N_{\mathrm{cp}}$ as the length of a whole OFDM block.

Let the transmitted signal pass through the above-described channel. The received signal in the $m$th block (after CP removal) at the BS without Doppler shifts compensation can be expressed as the following $1\!\times\! N$ vector
\begin{align} \label{ReceivedSignal}
{{\mathbf{y}}_{m}}=\sum\limits_{l=1}^{L}{\sum\limits_{p=1}^{P}{{{\rho }_{l,p}}\mathbf{a}^{T}\left( {{\theta }_{l,p}} \right){{\mathbf{S}}_{m}}\left( {{d}_{l}} \right){{\mathbf{\Phi }}_{m}}\left( \theta_{l,p} \right)}}+{{\mathbf{n}}_{m}},
\end{align}
where ${{\mathbf{\Phi }}_{m}}\left( \theta_{l,p} \right) = \operatorname{diag}\Big( \big[
   {{\beta }_{m,0}}\left( \theta_{l,p} \right), \  {{\beta }_{m,1}}\left( \theta_{l,p} \right),\  \ldots,$ $ {{\beta }_{m,N-1}}\!\left( \theta_{l,p} \right) \!
\big] \! \Big)$ with ${{\beta }_{m,n}}\!\left( \theta_{l,p} \right) \! = \! {{\mathrm{e}}^{\mathrm{j}2\mathrm{\pi }{{f}_{d}}\cos {{\theta }_{l,p}}\left( m{{N}_{s}}+n-{{d}_{l}} \right){{T}_{s}}}}$. Here, $T_s$ is the sampling interval. Besides, ${{\mathbf{n}}_{m}} \in \mathbb{C}^{1\times N}$ is the zero-mean complex additive white Gaussian noise (AWGN) in the $m$th block at the BS with $E\{{{\mathbf{n}}_{m}^H}{{\mathbf{n}}_{m}}\} \!=\! {\sigma_{\mathrm{n}}^{2}}{\mathbf{I}_N}$, where $\sigma_{\mathrm{n}}^{2}$ is the noise power.

Owing to the ambiguity between any two opposite directions about the ULA, it is sufficient to perform the multi-branch transmit beamforming towards a set of $Q$ directions ${{\vartheta }_{q}} \!\in\! \left( 0, \mathrm{\pi} \right), \!\ q\!=\!1,2,\ldots, Q$, which can cover the entire AoD range of $\left( 0, 2\mathrm{\pi} \right)$.
Then, the transmit beamforming and Doppler shifts compensation can be performed by substituting ${{\mathbf{S}}_{m}}\left( {{d}_{l}} \right)$ with ${{\mathbf{\tilde{S}}}_{m,q}}\left( {{d}_{l}} \right) \!=\! {{\mathbf{b}}^{*}}\left( {{\vartheta }_{q}} \right){{\mathbf{s}}_{m}}\left( {{d}_{l}} \right){{\mathbf{\Psi }}_{m,l}}\left( {{\vartheta }_{q}} \right)$, where $\mathbf{b}\left( {{\vartheta }_{q}} \right) \!=\! \frac{\eta}{M\sqrt{Q}}{{\mathbf{a}}}\left( {{\vartheta }_{q}} \right){{\mathrm{e}}^{\mathrm{j}\phi \left( {{\vartheta }_{q}} \right)}}$ represents the $q$th beamformer.
Here, $\phi \left( {{\vartheta }_{q}} \right)$ denotes the random phase introduced at $\mathbf{b}\left( {{\vartheta }_{q}} \right)$, and $\eta \!=\! \frac{1}{\left\| \sum\limits_{q=1}^{Q}{\frac{1}{M\sqrt{Q}}\mathbf{a}\left( {{\vartheta }_{q}} \right){{\mathrm{e}}^{\mathrm{j}\phi \left( {{\vartheta }_{q}} \right)}}} \right\|_2}$ is the normalization coefficient to keep the total transmit power per symbol to 1.
The associated Doppler shift compensation matrix is ${{\mathbf{\Psi }}_{m,l}}\left( {{\vartheta }_{q}} \right) \!=\! \operatorname{diag}\left( {{\left[ \begin{matrix}
   {{{\tilde{\beta }}}_{m,0,l}}\left( {{\vartheta }_{q}} \right),\!\! & {{{\tilde{\beta }}}_{m,1,l}}\left( {{\vartheta }_{q}} \right),\!\! & \ldots,\!\! & {{{\tilde{\beta }}}_{m,N-1,l}}\left( {{\vartheta }_{q}} \right)  \\
\end{matrix} \right]}^{T}} \right)$,
where ${{\tilde{\beta }}_{m,n,l}}\left( {{\vartheta }_{q}} \right) \!=\! {{\mathrm{e}}^{-\mathrm{j}2 {\mathrm{\pi}} {{f}_{d}}\cos {{\vartheta }_{q}}\left( m{{N}_{s}}+n-{{d}_{l}} \right){{T}_{s}}}}$.

Substituting ${{\mathbf{S}}_{m}}\left( {{d}_{l}} \right)$ with ${{\mathbf{\tilde{S}}}_{m,q}}\left( {{d}_{l}} \right)$ in (\ref{ReceivedSignal}), we arrive at
\begin{align} \label{Signal_Interference_Model}
 & {{\mathbf{r}}_{m,q}} =\sum\limits_{l=1}^{L}\sum\limits_{p=1}^{P}{{\rho }_{l,p}}\mathbf{a}^{T}\left( {{\theta }_{l,p}} \right){{\mathbf{b}}^{*}}\left( {{\vartheta }_{q}} \right){{\mathbf{s}}_{m}} \left( {{d}_{l}} \right) \nonumber \\
 & \hspace{23mm} \times  {{\mathbf{\Psi }}_{m,l}}\left( {{\vartheta }_{q}} \right){{\mathbf{\Phi }}_{m}}\left( \theta_{l,p} \right)+{{\mathbf{n}}_{m}}, \nonumber \\
 =& \underbrace{\frac{\eta}{\sqrt{Q}}{{\mathrm{e}}^{-\mathrm{j}\phi \left( {{\vartheta }_{q}} \right)}}\! \sum\limits_{l,p,{{\theta }_{l,p}}={{\vartheta }_{q}}} \! {{{\rho }_{l,p}}} {{\mathbf{s}}_{m}}\left( {{d}_{l}} \right)  }_{\mathrm{desired}\ \mathrm{signal}} \nonumber \\
 & + \! \underbrace{\sum\limits_{l,p,{{\theta }_{l,p}}\ne {{\vartheta }_{q}}} \! {{{\rho }_{l,p}}{{\mathbf{b}}^{H}}\! \left( {{\vartheta }_{q}} \right){{\mathbf{a}}}\left( {{\theta }_{l,p}} \right){{\mathbf{s}}_{m}}\left( {{d}_{l}} \right){{\mathbf{\Psi }}_{m,l}}\left( {{\vartheta }_{q}} \right){{\mathbf{\Phi }}_{m}}\left( \theta_{l,p} \right)}}_{\mathrm{interference}} \nonumber \\
 & + \! \underbrace{{{\mathbf{n}}_{m}}}_{\mathrm{noise}}.
\end{align}

When the number of antennas $M$ is massive, the interference in (\ref{Signal_Interference_Model}) tends to vanish, and the time-varying channel can be decomposed into a set of parallel time-invariant channels. However, the number of antennas may not be sufficiently large in practice, in which case there will still be uncompensated Doppler shifts due to limited spatial resolution while a thorough time-invariant equivalent channel cannot be achieved for each beamforming branch.
The Doppler spread\cite{Bellili_TC2017} could be employed here as a metric to measure the residual channel time variation.

The derivation of Doppler spread requires the channel PSD, which is the Fourier Transform of channel autocorrelation. Since different channel taps are independent and have identical statistical properties\cite{Norklit_TVT1999, Guo_TWC2019}, we only consider one tap for simplicity, i.e., $L\!=\!1, d_1\!=\!0$. By ignoring the noise item, the signal at the BS obtained after Doppler shifts compensation and multi-branch beamforming can be expressed as
\begin{align}{\label{ReceivedBeamSignal}}
{{\mathbf{r}}_{m}} & =\sum\limits_{q=1}^{Q}{{{\mathbf{r}}_{m,q}}} =\sum\limits_{q=1}^{Q}\sum\limits_{p=1}^{P}{{\rho }_{1,p}} \mathbf{a}^{T}\left( {{\theta }_{1,p}} \right){{\mathbf{b}}^{*}}\left( {{\vartheta }_{q}} \right) \nonumber \\
& \hspace{23mm} \times {{\mathbf{s}}_{m}}\left( {{d}_{1}} \right){{\mathbf{\Psi }}_{m,1}}\left( {{\vartheta }_{q}} \right){{\mathbf{\Phi }}_{m}}\left( \theta_{1,p} \right), \nonumber \\
& =\frac{\eta}{\sqrt{Q}} \sum\limits_{q=1}^{Q}\sum\limits_{p=1}^{P}{{\rho }_{p}}{{\mathrm{e}}^{-\mathrm{j}\phi \left( {{\vartheta }_{q}} \right)}} \frac{1}{M}\mathbf{a}^{H}\left( {{\vartheta }_{q}} \right){{\mathbf{a}}}\left( {{\theta }_{p}} \right) \nonumber \\
& \hspace{23mm} \times {{\mathbf{s}}_{m}}{{\mathbf{\Psi }}_{m,1}}\left( {{\vartheta }_{q}} \right){{\mathbf{\Phi }}_{m}}\left( \theta_p \right),
\end{align}
where $\rho_{1,p}$ and $\theta_{1,p}$ have been replaced by $\rho_{p}$ and $\theta_{p}$, respectively. Besides, the complex channel gain ${{\rho }_{p}}$ can be equivalently expressed as ${{\rho }_{p}} \!=\! {{\alpha }_{p}}{{\mathrm{e}}^{\mathrm{j}{{\varphi }_{p}}}}$, where ${{\alpha }_{p}} \!\sim\! \mathcal{N}\left( 0, 1/P \right)$ and ${{\varphi }_{p}} \!\sim\! U\left(0, {2\mathrm{\pi}}\right)$ denote the random channel gain and phase, respectively.

\section{Analysis of Channel PSD and Doppler Spread}

\subsection{Derivation of the Channel PSD}
The equivalent uplink channel of (\ref{ReceivedBeamSignal}) can be expressed in a continuous-time form as
\begin{align}{\label{ContinuousChannel}}
g\left( t \right) &=\frac{1}{\sqrt{Q}}\sum\limits_{q=1}^{Q} \int_{0}^{2\mathrm{\pi}} \alpha \left( \theta  \right) G\left( \cos \theta ,\cos{{\vartheta }_{q}} \right) \nonumber \\
& \hspace{18mm} \times {{\mathrm{e}}^{{\mathrm{j}} 2{\mathrm{\pi}} {{f}_{d}}t\cos \theta - {\mathrm{j}}2{\mathrm{\pi}} {{f}_{d}}t\cos {{\vartheta }_{q}} +\mathrm{j}\varphi \left( \theta  \right) -\mathrm{j}\phi \left( {{\vartheta }_{q}} \right)}}d\theta, \nonumber \\
& =\frac{1}{\sqrt{Q}}\sum\limits_{q=1}^{Q} \int_{0}^{2\mathrm{\pi}} \alpha \left( \theta  \right)G\left( \cos \theta ,\cos{{\vartheta }_{q}} \right) \nonumber \\
& \hspace{18mm} \times {{\mathrm{e}}^{\mathrm{j}{{\omega }_{d}}\left( \cos \theta -\cos {{\vartheta }_{q}} \right)t+\mathrm{j}\varphi \left( \theta  \right)-\mathrm{j}\phi \left( {{\vartheta }_{q}} \right)}}d\theta,
\end{align}
where ${{\omega }_{d}} \!=\! 2{\mathrm{\pi}} {{f}_{d}}$ and
$G\left( \cos \theta ,\cos{{\vartheta }_{q}} \right) \!=\! \frac{1}{M} \mathbf{a}^{H}\left( {{\vartheta }_{q}} \right)\mathbf{a}\left( \theta  \right) \!=\! \frac{1}{M} \sum\nolimits_{r=1}^{M}{{{\mathrm{e}}^{\mathrm{j2 }\chi (r-1) \left( \cos \theta -\cos {{\vartheta }_{q}} \right)}}}$.
Note that the equivalent continuous channel (\ref{ContinuousChannel}) is obtained by replacing $\theta_p$ with $\theta$, and $\alpha(\theta)$ and $\varphi(\theta)$ denote the random gain and phase for the path with AoD $\theta$. The normalization coefficient $\eta$ is omitted in the continuous-form channel (\ref{ContinuousChannel}) for simplicity, since it does not affect the following PSD analysis. Besides, $| G\left( \cos \theta ,\cos{{\vartheta }_{q}} \right)\! |^{2}$ is in fact the radiation pattern at direction $\theta$ with $\frac{1}{M} \mathbf{a}\left( {{\vartheta }_{q}} \right)$ as beamformer. Moreover, by fixing ${\vartheta }_{q} \!=\! \frac{\mathrm{\pi}}{2} $ and varying $\theta$, $| G\left( \cos \theta ,\cos{{\vartheta }_{q}} \right)\! |^{2}$ is exactly the radiation pattern obtained with the MF beamformer pointing towards the normal direction of ULA.

The autocorrelation for the equivalent continuous channel $g\left( t \right) $ is given by
\begin{align} {\label{ChannelAutocorrelation}}
& {{R}_{g}}\left( \tau  \right) =  E\left\{ g\left( t \right){{g}^{*}}\left( t+\tau  \right) \right\}, \nonumber \\
 = \ \! & \frac{1}{Q}\sum\limits_{q=1}^{Q} \sum\limits_{k=1}^{Q} \int_{0}^{2\mathrm{\pi}} \int_{0}^{2\mathrm{\pi}} E\Big\{ \alpha \left( \theta  \right){{\alpha }^{*}}( {\tilde{\theta }} ){{\mathrm{e}}^{\mathrm{j}\left[ \varphi \left( \theta  \right)-\varphi ( {\tilde{\theta }} ) \right]}} \nonumber \\
 & \hspace{2.3em} \times {{\mathrm{e}}^{\mathrm{j}\left[ \phi \left( {{\vartheta }_{k}} \right)-\phi \left( {{\vartheta }_{q}} \right) \right]}} G\left( \cos \theta ,\cos{{\vartheta }_{q}} \right) {{G}^{*}}\big( \cos \tilde{\theta },\cos{{\vartheta }_{k}} \big) \nonumber \\
 & \hspace{2.3em} \times  {{\mathrm{e}}^{\mathrm{j}{{\omega }_{d}}\left( \cos \theta -\cos {{\vartheta }_{q}} \right)t-\mathrm{j}{{\omega }_{d}}\left( \cos \tilde{\theta }-\cos {{\vartheta }_{k}} \right)\left( t+\tau  \right)}} \Big\} d\theta d\tilde{\theta }, \nonumber \\
 \overset{*}{\mathop{=}}\, \! \ \! & \frac{1}{Q}\sum\limits_{q=1}^{Q} \int_{0}^{2\mathrm{\pi}} E \big\{ {{\left| \alpha \left( \theta  \right) \right|}^{2}} \big\}  {{\left| G\left( \cos \theta ,\cos{{\vartheta }_{q}} \right) \right|}^{\mathrm{2}}} \nonumber \\
 & \hspace{2.3em} \times {{\mathrm{e}}^{-\mathrm{j}{{\omega }_{d}}\left( \cos \theta -\cos {{\vartheta }_{q}} \right)\tau }}d\theta, \nonumber \\
\overset{**}{\mathop{=}}\, \! \ \! &\frac{1}{ 2\mathrm{\pi} Q} \sum\limits_{q=1}^{Q}{\int_{0}^{2\mathrm{\pi}} {{{\left| G\left( \cos \theta ,\cos{{\vartheta }_{q}} \right) \right|}^{\mathrm{2}}}{{\mathrm{e}}^{-\mathrm{j}{{\omega }_{d}}\left( \cos \theta -\cos {{\vartheta }_{q}} \right)\tau }}d\theta }}, \nonumber \\
= \ \! &\frac{1}{ \mathrm{\pi} Q} \sum\limits_{q=1}^{Q}{\int_{0}^{\mathrm{\pi}} {{{\left| G\left( \cos \theta ,\cos{{\vartheta }_{q}} \right) \right|}^{\mathrm{2}}}{{\mathrm{e}}^{-\mathrm{j}{{\omega }_{d}}\left( \cos \theta -\cos {{\vartheta }_{q}} \right)\tau }}d\theta }},
\end{align}
where $\overset{*}{\mathop{=}}\, \!$ employs the properties
\begin{align*}
E\Big\{ {{\mathrm{e}}^{\mathrm{j}\left[ \varphi \left( \theta  \right)-\varphi ( {\tilde{\theta }} ) \right]}} \Big\} & =\left\{ \begin{matrix}
   1,\ \theta =\tilde{\theta }  \\
   0,\ \theta \ne \tilde{\theta }  \\
\end{matrix} \right.,\ \\
E\Big\{ {{\mathrm{e}}^{\mathrm{j}\left[ \phi \left( {{\vartheta }_{k}} \right)-\phi \left( {{\vartheta }_{q}} \right) \right]}} \Big\} & =\left\{ \begin{matrix}
   1,\ q=k  \\
   0,\ q\ne k  \\
\end{matrix} \right.,
\end{align*}
and $\overset{**}{\mathop{=}}\, \!$ comes from $\int_{0}^{2\mathrm{\pi}} {{{\left| \alpha \left( \theta  \right) \right|}^{2}}d\theta } \!=\! 1$ and $E \big\{ {{\left| \alpha \left( \theta  \right) \right|}^{2}} \big\} \!=\! \frac{1}{2\mathrm{\pi}}$.

The channel PSD is the Fourier transform of the channel autocorrelation ${{R}_{g}}\left( \tau  \right)$ and the explicit expression of channel PSD is provided by the following \emph{Lemma}.

\begin{lemma} {\label{Lemma1}}
Let $\omega$ be the Doppler frequency and denote $\tilde{\omega} \!=\! \frac{\omega}{\omega_d} \!=\! \frac{\omega}{2{\mathrm{\pi}}f_d}$ as the normalized Doppler frequency with respect to the maximum Doppler shift. Then, for the given channel autocorrelation $R_g(\tau)$ in (\ref{ChannelAutocorrelation}), the channel PSD can be expressed in the form of
\begin{align}{\label{PSD}}
P\left( \omega  \right) = \frac{1}{{{\omega }_{d}}}  {{\left| \mathcal{G}(\tilde{\omega}) \right|}^{2}} \mathcal{W}\left( {\tilde{\omega }} \right),
\end{align}
where
\begin{align}{\label{PatternFunction}}
| \mathcal{G}(\tilde{\omega})|^{2} = | \frac{1}{M} \sum\nolimits_{r=1}^{M} { {{\mathrm{e}}^{-\mathrm{j2} \chi (r-1) \tilde{\omega }}}} |^{2} = \frac{{{\sin }^{2}}\left( \chi M \tilde{\omega } \right)}{M^2 {{\sin }^{2}}\left( \chi \tilde{\omega } \right)},
\end{align}
and
\begin{align}{\label{DistortionFunction}}
\mathcal{W}\left( {\tilde{\omega }} \right) = \frac{2}{ Q} \sum\limits_{q=1}^{Q}{\frac{1}{\sqrt{1-{{\left( \tilde{\omega }-\cos {{\vartheta }_{q}} \right)}^{2}}}}{{\mathcal{I}}_{q}}\left( {\tilde{\omega }} \right)},
\end{align}
are named as pattern function and beam-distortion function, respectively.
Note that ${{\mathcal{I}}_{q}}\left( {\tilde{\omega }} \right)$ is the binary-value indicator function defined in the proof below.

\begin{IEEEproof}
According to the definition, the channel PSD can be expressed as
\begin{align}{\label{PSDDefinition}}
 P\left( \omega  \right) & = \int_{-\infty }^{+\infty }{{{R}_{g}}\left( \tau  \right){{\mathrm{e}}^{-\mathrm{j}\omega \tau }}d\tau }, \nonumber \\
 & = \frac{1}{ \mathrm{\pi} Q} \sum\limits_{q=1}^{Q}\int_{0}^{\mathrm{\pi}} {{\left| G\left( \cos \theta ,\cos{{\vartheta }_{q}} \right) \right|}^{\mathrm{2}}} \nonumber \\
 & \hspace{4.5em} \times \left[ \int_{-\infty }^{+\infty }{{{\mathrm{e}}^{-\mathrm{j}{{\omega }_{d}} \left( \cos \theta -\cos {{\vartheta }_{q}} \right)\tau }} {{\mathrm{e}}^{-\mathrm{j}\omega \tau }}d\tau } \right] \! d\theta, \nonumber \\
 & = \frac{2}{Q} \sum\limits_{q=1}^{Q} \int_{0}^{\mathrm{\pi}} {{\left| G\left( \cos \theta ,\cos{{\vartheta }_{q}} \right) \right|}^{\mathrm{2}}}  \nonumber \\
 & \hspace{4.5em} \times  \delta \left( \omega +{{\omega }_{d}} \left( \cos \theta -\cos {{\vartheta }_{q}} \right) \right) d\theta,
\end{align}
where we have exploited
$\int_{-\infty }^{+\infty }{{{\mathrm{e}}^{-\mathrm{j}{{\omega }_{d}}\left( \cos \theta -\cos {{\vartheta }_{q}} \right)\tau }}{{\mathrm{e}}^{-\mathrm{j}\omega \tau }}d\tau } \!=\! \mathrm{2 }{\mathrm{\pi}} \delta \left( \omega \!+\! {{\omega }_{d}}\left( \cos \theta \!-\! \cos {{\vartheta }_{q}} \right) \right)$.

In addition, there holds
\begin{align} {\label{delta_Integral}}
  & \int_{0}^{\mathrm{\pi}} {{{\left| G\left( \cos \theta ,\cos{{\vartheta }_{q}} \right) \right|}^{\mathrm{2}}}\delta \left( \omega +{{\omega }_{d}}\left( \cos \theta -\cos {{\vartheta }_{q}} \right) \right)d\theta } \nonumber \\
&  \hspace{-5mm} \overset{y={{\omega }_{d}}\cos \theta }{\mathop{=}}\,\frac{1}{{{\omega }_{d}}} \int_{-{{\omega }_{d}} }^{{{\omega }_{d}} }  {{\left| G\left( \frac{y} {{{\omega }_{d}}}, \cos{{\vartheta }_{q}} \right) \right|}^{\mathrm{2}}} \frac{1} {\sqrt{1 -{{\left( \frac{y} {{{\omega }_{d}}} \right)}^{2}}}} \nonumber \\
& \hspace{20mm} \times \delta \left( y+\omega -{{\omega }_{d}}\cos {{\vartheta }_{q}} \right)dy, \nonumber \\
 =& \left\{ \begin{matrix}
   \frac{1}{{{\omega }_{d}}}{{\left| G\left( \cos {{\vartheta }_{q}} \!-\! \frac{\omega }{{{\omega }_{d}}}, \cos {{\vartheta }_{q}} \right) \right|}^{2}} \frac{1}{\sqrt{1- {{\left( \frac{\omega }{{{\omega }_{d}}} -\cos {{\vartheta }_{q}} \right)}^{2}}}}, \\
   \hspace{10.5em}  -1 \!\le\! \frac{\omega }{{{\omega }_{d}}} \!-\! \cos {{\vartheta }_{q}} \!\le\! 1  \\
   0, \hspace{12.2em} \mathrm{otherwise} \hspace{1.8em}
\end{matrix} \right.\!, \nonumber \\
 =& \left\{ \begin{matrix}
   \frac{1}{{{\omega }_{d}}} {{\left| \mathcal{G}(\tilde{\omega}) \right|}^{2}} \frac{1}{\sqrt{1-{{\left( \tilde{\omega }-\cos {{\vartheta }_{q}} \right)}^{2}}}}, -1 \!\le\! {\tilde{\omega} } \!-\! \cos {{\vartheta }_{q}} \!\le\! 1  \\
   0,  \hspace{9.2em} \mathrm{otherwise} \hspace{3.5em}
\end{matrix} \right..
\end{align}

Combining (\ref{PSDDefinition}) and (\ref{delta_Integral}), we obtain
\begin{align}
 P\left( {\omega } \right) & = \frac{2}{ Q} \sum\limits_{q=1}^{Q}{\frac{1}{{{\omega }_{d}}} {{\left| \mathcal{G}(\tilde{\omega}) \right|}^{2}} \frac{1}{\sqrt{1-{{\left( \tilde{\omega }-\cos {{\vartheta }_{q}} \right)}^{2}}}} {{\mathcal{I}}_{q}} \left( {\tilde{\omega }} \right)} \nonumber \\
& =\frac{1}{{{\omega }_{d}}}  {{\left| \mathcal{G}(\tilde{\omega}) \right|}^{2}}   W\left( {\tilde{\omega }} \right).
\end{align}
Here, ${{\mathcal{I}}_{q}}\left( {\tilde{\omega }} \right) \!=\! \left\{ \begin{matrix}
   1,\ q\in \mathcal{S}\left( {\tilde{\omega }} \right)  \\
   0,\ q\notin \mathcal{S}\left( {\tilde{\omega }} \right)  \\
\end{matrix} \right.$, with $\mathcal{S}\left( {\tilde{\omega }} \right)$ being the set of beamforming branches contributing to the PSD at $\tilde{\omega}$. From the derivation (\ref{delta_Integral}), $\mathcal{S}\left( {\tilde{\omega }} \right)$ can be given by $\mathcal{S}\left( {\tilde{\omega }} \right) \!=\! \{ q \ \! | \ \! \tilde{\omega } \!-\! 1 \!\le\! \cos {{\vartheta }_{q}} \!\le\! \tilde{\omega } \!+\! 1 \}$. However, there is also an implicit constraint about ${{\vartheta }_{q}} \in (0, \mathrm{\pi})$, i.e., $ -1 \!\le\! \cos {{\vartheta }_{q}} \!\le\! 1 $. By making the implicit constraint explicit, $\mathcal{S}\left( {\tilde{\omega }} \right)$ can be re-expressed as
\begin{align} {\label{Somega}}
 \mathcal{S}\left( {\tilde{\omega }} \right) & \!=\! \left\{ q \ \! | \ \! \tilde{\omega } \!-\! 1 \!\le\! \cos {{\vartheta }_{q}}\le \tilde{\omega } \!+\! 1,\ -1 \!\le\! \cos {{\vartheta }_{q}} \!\le\! 1 \right\} \nonumber \\
& \!=\! \left\{ \begin{matrix}
   \left\{ q \ \! | -1 \!\le\! \cos {{\vartheta }_{q}} \!\le\! \tilde{\omega } \!+\! 1 \right\},\ -2 \!\le\! \tilde{\omega } \!<\! 0  \\
   \left\{ q \ \! | \ \! \tilde{\omega } \!-\! 1 \!\le\! \cos {{\vartheta }_{q}} \!\le\! 1 \right\},\ \ \ \ \ \ 0 \!\le\! \tilde{\omega } \!\le\! 2 \\
\end{matrix} \right..
\end{align}

This completes the proof.
\end{IEEEproof}
\end{lemma}

From \emph{Lemma 1}, the following observations can be made:

1) The expression of $\mathcal{S}\left( {\tilde{\omega }} \right)$ in (\ref{Somega}) reveals that the PSD is nonzero only for $\left| {\tilde{\omega }} \right| \!\le\! 2$. Obviously, the maximum Doppler frequency $\omega_{\mathrm{max}}$ is $\left| \omega_{\mathrm{max}} \right| = 2\ \! \omega_d$.

2) The most interesting observation from (\ref{PSD}) is that the channel PSD can be fully characterized by $| \mathcal{G}(\tilde{\omega}) |^{2}$ and $\mathcal{W}(\tilde{\omega})$.
Taking ${{\vartheta }_{q}} \!=\! \frac{\mathrm{\pi}}{2}$ and $ -\tilde{\omega} \!=\! \cos \theta \!-\! \cos {{\vartheta }_{q}} \!=\! \cos \theta$, we arrive at ${\mathcal{G}} \left( \tilde{\omega} \right) \!=\! G\left( \cos \theta ,\cos \frac{\mathrm{\pi}}{2} \right)$, which implies that $| {\mathcal{G}} \left( \tilde{\omega} \right) |^{2} $ is the converted radiation pattern obtained with the MF beamformer pointing to the normal direction of ULA. This explains why $| \mathcal{G}(\tilde{\omega})|^{2}$ is named as pattern function.
Besides, ${{\mathcal{I}}_{q}}\left( {\tilde{\omega }} \right)$ is the binary-value indicator function indicating whether the $q$th beamforming branch contributes to the PSD at $\tilde{\omega}$. Therefore, $\mathcal{W}\left( {\tilde{\omega }} \right) $ reflects the comprehensive impact of different beamformers on channel PSD, and is named as beam-distortion function to highlight its distortion effect on the pattern function $| \mathcal{G}(\tilde{\omega})|^{2}$.
Moreover, the pattern function $| \mathcal{G}(\tilde{\omega}) |^{2}$ only depends on the antenna spacing $d$, and the beam-distortion function $\mathcal{W}(\tilde{\omega})$ is entirely determined by the configuration of beamforming directions $\vartheta_q, q\!=\! 1,2,\ldots, Q $.

3) The PSD in (\ref{PSD}) can be equivalently written as $P\left( \omega \right) \!=\! \frac{1}{{{\omega }_{d}}}  {{\left| \mathcal{G}\left( \frac{\omega }{\omega_d} \right) \right|}^{2}} \mathcal{W}\left( \frac{\omega }{\omega_d} \right)$. Evidently, increasing $\omega_d$, i.e., the maximum Doppler shift $f_d$, will preserve the shape of the PSD, except that the resulting PSD will be linearly stretched in frequency and reversely decreased in amplitude. Nevertheless, the integral of $P\left( \omega \right)$ with respect to $\omega$ is independent of $\omega_d$, because of
\begin{align*}
\int_{-2\ \!\!\omega_d}^{2\ \!\!\omega_d} {P\left( \omega \right) d\omega } &= \int_{-2\ \!\!\omega_d}^{2\ \!\! \omega_d} { {{\left| \mathcal{G}\left( \frac{\omega }{\omega_d} \right) \right|}^{2}} \mathcal{W}\left( \frac{\omega }{\omega_d} \right) d\frac{\omega }{\omega_d} } \nonumber \\
& = \int_{-2 }^{2} { {{\left| \mathcal{G}(\tilde{\omega}) \right|}^{2}} \mathcal{W}\left( {\tilde{\omega }} \right) d\tilde{\omega} }.
\end{align*}

4) The Doppler spread can be calculated as
\begin{align} {\label{DopplerSpread}}
{{\sigma }_{\mathrm{DS}}} & =\sqrt{\frac{\int_{-2\ \!\!{{\omega }_{d}}}^{2\ \!\!{{\omega }_{d}}}{{{\omega }^{2}} P\left( \omega  \right)d\omega }} {\int_{-2\ \!\!{{\omega }_{d}}}^{2\ \!\! {{\omega }_{d}}}{P\left( \omega  \right)d\omega }}} \nonumber \\
& \hspace{-2mm} \overset{\omega ={{\omega }_{d}}\tilde{\omega }}{\mathop{=}}\, \  {{\omega }_{d}}\sqrt{\frac{\int_{-2}^{2 }{{{{\tilde{\omega }}}^{2}}{{\left| \mathcal{G} \left( {\tilde{\omega }} \right) \right|}^{2}} \mathcal{W}\left( {\tilde{\omega }} \right)d\tilde{\omega }}}{\int_{-2}^{2} {{{\left| \mathcal{G} \left( {\tilde{\omega }} \right) \right|}^{2}} \mathcal{W}\left( {\tilde{\omega }} \right)d\tilde{\omega }}}}.
\end{align}
Considering that the two integrals with respect to $\tilde{\omega}$ in (\ref{DopplerSpread}) does not depend on $\omega_d$, we know that the Doppler spread ${{\sigma }_{\mathrm{DS}}}$ is linearly proportional to $\omega_d$, i.e., the maximum Doppler shift $f_d$. In other words, the higher the HSR velocity is, the larger the Doppler spread ${{\sigma }_{\mathrm{DS}}}$ will be.

\begin{remark}
The derivation of channel PSD can be readily extended to much more generalized cases. Consider that a linear antenna array (possibly non-uniform) is equipped at the HSR and denote $\Delta {{d}_{r}}$ as the antenna spacing between the $r$th antenna and the first one, with $\Delta {{d}_{1}} = 0$.
Moreover, we assume that the signal AoDs $\theta \!\sim\! U\left(\theta_{\mathrm{L}}, \theta_{\mathrm{R}} \right)$, where $\theta_{\mathrm{L}}, \theta_{\mathrm{R}} $ denote the bounds of the AoD region. Similar to~\cite{You_TWC2015}, we denote $\kappa(\theta)$ as the complex-valued channel gain corresponding to the AoD $\theta$.
The channels with different AoDs are assumed uncorrelated, i.e., $E\{ \kappa(\theta) \kappa^*(\theta') \} \!=\! \rho(\theta) \delta(\theta-\theta')$, where $\rho(\theta)$ represents the channel power angle spectrum (PAS) which models the channel power distribution in angle domain~\cite{You_TWC2015}.
There holds $\int_{\theta_{\mathrm{L}}}^{\theta_{\mathrm{R}}} \rho(\theta) d\theta\!=\! 1$ such that the total channel gain is normalized to 1.

By introducing the concept of channel PAS $\rho\left( \theta  \right)$, our PSD derivation can cover a wide variety of channel circumstances.
For example, if the channel has uniform PAS in non-line-of-sight (NLoS) environments, there is $\rho(\theta) \!=\! \frac{1}{ {{{\theta }_{\mathrm{R}}} - {{\theta }_{\mathrm{L}}}}}, \ \theta_{\mathrm{L}} \!\le\! \theta \!\le\! \theta_{\mathrm{R}}$. Instead, in the case of line-of-sight (LoS) environments with LoS path component at $\theta_{\mathrm{LoS}}$ and NLoS subpaths, we have
\begin{align*}
\rho(\theta) = \rho_{\mathrm{NLoS}}(\theta) + \frac{K}{K\!+\!1} \delta\left( \theta \!-\! \theta_{\mathrm{LoS}} \right), \ \theta_{\mathrm{L}} \!\le\! \theta \!\le\! \theta_{\mathrm{R}},
\end{align*}
where $\rho_{\mathrm{NLoS}}(\theta)$ denotes the channel PAS for NLoS subpaths, with $\int_{\theta_\mathrm{L}}^{\theta_\mathrm{R}} \rho_{\mathrm{NLoS}}\left( \theta  \right) d\theta = \frac{1}{K+1}$. Here, $K$ is the Rician factor to reflect the power ratio between LoS component and NLoS subpaths.

Following the similar derivation as in \emph{Lemma 1}, the pattern function and beam-distortion function in the above-described generalized scenario turn to be
\begin{align}
| \mathcal{G}(\tilde{\omega})|^{2} = \left| \frac{1}{M} \sum\limits_{r=1}^{M} {{{\mathrm{e}}^{-\mathrm{j2 \pi} \frac{\Delta {{d}_{r}}}{\lambda } \tilde{\omega }}}} \right|^{2},
\end{align}
and
\begin{align}
\mathcal{W}\left( {\tilde{\omega }} \right) = \frac{2\mathrm{\pi}}{Q} \sum\limits_{q=1}^{Q}{\frac{ \rho\left( \arccos \left( \cos {{\vartheta }_{q}} \!-\! \tilde{\omega } \right) \right) }{\sqrt{1-{{\left( \tilde{\omega }-\cos {{\vartheta }_{q}} \right)}^{2}}}}{{\mathcal{I}}_{q}}\left( {\tilde{\omega }} \right)},
\end{align}
where ${{\mathcal{I}}_{q}}\left( {\tilde{\omega }} \right) \!=\! \left\{ \begin{matrix}
   1,\ q\in \mathcal{S}\left( {\tilde{\omega }} \right)  \\
   0,\ q\notin \mathcal{S}\left( {\tilde{\omega }} \right)  \\
\end{matrix} \right.$ remains the same while $\mathcal{S}\left( {\tilde{\omega }} \right)$ becomes
\begin{align}
& \mathcal{S}\left( {\tilde{\omega }} \right) \!=\! \{ q \ \! | \ \! \tilde{\omega } \!+\! \cos {{\theta }_{\mathrm{R}}} \!\le\! \cos {{\vartheta }_{q}}\le \tilde{\omega } \!+\! \cos {{\theta }_{\mathrm{L}}}, \nonumber \\
& \hspace{22mm} \cos {{\theta }_{\mathrm{R}}} \!\le\! \cos {{\vartheta }_{q}} \!\le\! \cos {{\theta }_{\mathrm{L}}} \} \nonumber \\
= & \left\{ \begin{matrix}
   \left\{ q \ \! | \cos {{\theta }_{\mathrm{R}}} \!\le\! \cos {{\vartheta }_{q}} \!\le\! \tilde{\omega } \!+\! \cos {{\theta }_{\mathrm{L}}} \right\},\ -\mu \left( {{\theta }_{\mathrm{L}}},{{\theta }_{\mathrm{R}}} \right) \!\le\! \tilde{\omega } \!<\! 0  \\
   \left\{ q \ \! | \ \! \tilde{\omega } \!+\! \cos {{\theta }_{\mathrm{R}}} \!\le\! \cos {{\vartheta }_{q}} \!\le\! \cos {{\theta }_{\mathrm{L}}} \right\},\ \ \ 0 \!\le\! \tilde{\omega } \!\le\! \mu\left( {{\theta }_{\mathrm{L}}},{{\theta }_{\mathrm{R}}} \right) \\
\end{matrix} \right..
\end{align}
Note that $\mu \left( {{\theta }_{\mathrm{L}}},{{\theta }_{\mathrm{R}}} \right) \!=\! \cos {{\theta }_{\mathrm{L}}} \!-\! \cos {{\theta }_{\mathrm{R}}}$.
In such case, the pattern function $| \mathcal{G}(\tilde{\omega}) |^{2}$ depends on the antenna spacings $\Delta d_r$, and the beam-distortion function $\mathcal{W}(\tilde{\omega})$ is jointly determined by the AoD region $\left( {{\theta }_{\mathrm{L}}},{{\theta }_{\mathrm{R}}} \right)$, the channel PAS $\rho(\theta)$ and the configuration of beamforming directions $\vartheta_q, \!\ q\!=\! 1,2,\ldots, Q $.

It is confirmed that the derived PSD can be extended to non-uniform linear arrays and non-Jakes' channels with generic channel PAS. Nonetheless, unless otherwise specified, we will limit the discussion hereinbelow in the scope of ULA and Jakes' channel for simplicity.
\end{remark}

\subsection{Impact of Beamforming Directions and Antenna Spacing on Channel PSD}
In this section, we discuss how the configuration of beamforming directions and the choice of antenna spacing influence the channel PSD.

\subsubsection{Impact of beamforming directions}
As the number of selected beamformers $Q$ goes to infinity, the beam-distortion function given in (\ref{DistortionFunction}) can be transformed into the following integral form
\begin{align}
\mathcal{W}\left( {\tilde{\omega }} \right) & = \frac{2}{Q}\sum\limits_{q=1}^{Q}{\frac{1}{\sqrt{1 \!-\! {{\left( \tilde{\omega } \!-\! \cos {{\vartheta }_{q}} \right)}^{2}}}}{{\mathcal{I}}_{q}}\left( {\tilde{\omega }} \right)} \nonumber \\
& = 2 \int_{0}^{\mathrm{\pi}} {\frac{1}{\sqrt{1 \!-\! {{\left( \tilde{\omega } \!-\! \cos \vartheta  \right)}^{2}}}}f\left( \vartheta  \right)\mathcal{I}\left( \vartheta ,\tilde{\omega } \right)d\vartheta },
\end{align}
where $\vartheta$ is the continuous counterpart of $\vartheta_q$, $f(\vartheta)$ is the probability density function (pdf) of $\vartheta$, and ${{\mathcal{I}}}\left( \vartheta, {\tilde{\omega }} \right) \!=\! \left\{ \begin{matrix}
   1,\ \vartheta \in \mathcal{S}\left( \vartheta, {\tilde{\omega }} \right)  \\
   0,\ \vartheta \notin \mathcal{S}\left( \vartheta, {\tilde{\omega }} \right)  \\
\end{matrix} \right.$ is the binary-value indicator function, with
\begin{align}
 \mathcal{S}\left( \vartheta ,\tilde{\omega } \right) & \!=\! \left\{ \begin{matrix}
   \left\{ \vartheta \ \! | -1 \!\le\! \cos \vartheta \!\le\! \tilde{\omega } \!+\! 1 \right\},\ -2 \!\le\! \tilde{\omega } \!<\! 0  \\
   \left\{ \vartheta \ \! |\ \! \tilde{\omega } \!-\! 1 \!\le\! \cos \vartheta \!\le\! 1 \right\}, \ \ \ \ \ \  0 \!\le\! \tilde{\omega } \!\le\! 2 \\
\end{matrix} \right. \nonumber \\
 & \!=\! \left\{ \begin{matrix}
   \left\{ \vartheta \ \! | \arccos \left( \tilde{\omega } \!+\! 1 \right) \!\le\! \vartheta \!\le\! \mathrm{\pi} \right\},\ -2 \!\le\! \tilde{\omega } \!<\! 0  \\
   \left\{ \vartheta \ \! | \ \! 0 \!\le\! \vartheta \!\le\! \arccos \left( \tilde{\omega } \!-\! 1 \right) \right\},\ \ \ 0\!\le\! \tilde{\omega } \!\le\! 2 \\
\end{matrix} \right..
\end{align}
Note that $\mathcal{S}\left( \vartheta ,\tilde{\omega } \right)$ can be directly derived from $\mathcal{S}\left( \tilde{\omega } \right)$, by substituting $\vartheta_q$ in (\ref{Somega}) with $\vartheta$.

Next, we further derive a more explicit form of the beam-distortion functions, under two typical configurations of beamforming directions: First, the beamforming directions are configured such that $\cos \vartheta_q, \!\ q\!=\! 1,2,\ldots, Q $ are evenly distributed between $ ( -1, \ \! 1 )$; second, the beamforming directions $ \vartheta_q, q\!=\! 1,2,\ldots, Q $ themselves are evenly configured between $ \left(0, \mathrm{\pi} \right)$. We refer to the two configurations of beamforming directions as ``Equi-cos'' and ``Equi-angle'', respectively. Note that ``Equi-cos'' is considered since the multi-branch beamforming with such configured beamformers can be implemented efficiently with fast Fourier transform (FFT)\cite{Zhang_TWC2018}.

Case 1: In the case of ``Equi-cos'', i.e., $\cos \vartheta_q, q\!=\! 1,2,\ldots, Q $ are evenly distributed between $ ( -1, \ \! 1 )$,
the pdf of $\vartheta$ can be expressed as
\begin{align} {\label{DensityEquiCos}}
f\left( \vartheta \right)=\frac{ 1}{2} \sin \vartheta, \ \vartheta \sim (0, \mathrm{\pi}).
\end{align}
Here, the pdf (\ref{DensityEquiCos}) should be in sinusoidal form since ``Equi-cos'' distribution implies $-d\cos\vartheta \!=\! \sin\vartheta d\vartheta$, and the normalization term $\frac{ 1}{2}$ comes from $\int_{0}^{\mathrm{\pi}} {\sin \vartheta d\vartheta } \!=\! 2$.

As a result, the beam-distortion function can be expressed as
\begin{align} {\label{DistortionFunction_EquiCos0}}
\mathcal{W}\left( {\tilde{\omega }} \right)  &= {{\left. \frac{2}{Q}\sum\limits_{q=1}^{Q} {\frac{1}{\sqrt{1-{{\left( \tilde{\omega }-\cos {{\vartheta }_{q}} \right)}^{2}}}} {\mathcal{I}}_{q}\left( {\tilde{\omega }} \right)} \right|}_{ \cos {{\vartheta }_{q}} \ \! \sim \ \! \mathcal{U}(-1,1) }} \nonumber \\
 & = \int_{0}^{ \mathrm{\pi} } {\frac{\sin \vartheta }{\sqrt{1-{{\left( \cos \vartheta -\tilde{\omega } \right)}^{2}}}} \mathcal{I}\left( \vartheta ,\tilde{\omega } \right)d\vartheta }.
\end{align}

To further simplify (\ref{DistortionFunction_EquiCos0}), we take a variable substitution of $x= \arccos\left( \cos \vartheta - \tilde{\omega } \right)$, i.e., $\vartheta = \arccos(\cos x + \tilde{\omega}) $. Then, the indicator function $\mathcal{I}\left( \vartheta ,\tilde{\omega } \right)$ becomes $\mathcal{I}\left( \arccos(\cos x + \tilde{\omega}),\tilde{\omega } \right) =  {{\left. {{\mathcal{I}}}\left( \vartheta, {\tilde{\omega }} \right) \right|}_{\vartheta = \arccos(\cos x + \tilde{\omega})}} $, with the beamformer set $\mathcal{S}\left( \vartheta ,\tilde{\omega } \right)$ being transformed into
\begin{align}
& \mathcal{S} \left( \arccos(\cos x \!+\! \tilde{\omega}),\tilde{\omega } \right) \nonumber \\
=& \left\{ \begin{matrix}
   \left\{ x \ \! | \ \! 0 \!\le\! x \!\le\! \arccos \left( -1 \!-\! \tilde{\omega } \right)  \right\}, -2 \!\le\! \tilde{\omega } \!<\! 0  \\
   \left\{ x \ \! | \arccos \left( 1 \!-\! \tilde{\omega } \right) \!\le\! x \!\le\! \mathrm{\pi} \right\}, \ \ \ \ 0 \!\le\! \tilde{\omega } \!\le\! 2 \\
\end{matrix} \right..
\end{align}
After the variable substitution, (\ref{DistortionFunction_EquiCos0}) can be finally expressed in a closed form as
\begin{align} {\label{DistortionFunction_EquiCos}}
\mathcal{W}\left( {\tilde{\omega }} \right) &= \int_{\arccos\left( 1-\tilde{\omega } \right)}^{\arccos\left( -1-\tilde{\omega } \right)}{\mathcal{I}\left( \arccos(\cos x + \tilde{\omega}),\tilde{\omega } \right)dx}, \nonumber \\
 & = \left\{ \begin{matrix}
 \int_{0}^{\arccos \left( -1-\tilde{\omega } \right)}{1dx},\ -2 \le \tilde{\omega }<0  \\
 \int_{\arccos \left( 1-\tilde{\omega } \right)}^{ \mathrm{\pi} }{1dx},\ \ \ \ \ 0\le \tilde{\omega }\le 2 \\
\end{matrix} \right., \nonumber \\
 & = \arccos \left( \left| {\tilde{\omega }} \right| \!-\! 1 \right), \ \! \left| {\tilde{\omega }} \right| \le 2 \ \!.
\end{align}

Case 2: In the case of ``Equi-angle'', i.e., $ \vartheta_q, q\!=\!1,2,\ldots, Q $ are evenly selected between $ \left(0, \mathrm{\pi} \right)$, the pdf of $\vartheta$ can be given by
\begin{align}
f\left( \vartheta  \right)=\frac{1}{\mathrm{\pi}}, \ \vartheta \sim (0, \mathrm{\pi}).
\end{align}
As a result, the beam-distortion function can be expressed as
\begin{align} {\label{DistortionFunction_EquiAngle}}
\mathcal{W}\left( {\tilde{\omega }} \right) &= {{\left. \frac{2}{Q}\sum\limits_{q=1}^{Q} {\frac{1}{\sqrt{1-{{\left( \tilde{\omega }-\cos {{\vartheta }_{q}} \right)}^{2}}}} {\mathcal{I}}_{q}\left( {\tilde{\omega }} \right)} \right|}_{ {{\vartheta }_{q}} \ \! \sim \ \! \mathcal{U}(0,\mathrm{\pi}) }}, \nonumber \\
& = \frac{2}{ \mathrm{\pi} } \int_{0}^{\mathrm{\pi}} { \frac{1}{\sqrt{1-{{\left( \cos \vartheta -\tilde{\omega } \right)}^{2}}}}\mathcal{I}\left( \vartheta ,\tilde{\omega } \right)d\vartheta }, \nonumber \\
 & =\left\{ \begin{matrix}
   \frac{2}{ \mathrm{\pi} } \int_{\arccos \left(\tilde{\omega } + 1 \right)}^{ \mathrm{\pi} }{\frac{1}{\sqrt{1-{{\left( \cos \vartheta -\tilde{\omega } \right)}^{2}}}}d\vartheta },\ -2 \le \tilde{\omega }<0  \\
   \frac{2}{ \mathrm{\pi} } \int_{0}^{\arccos \left(\tilde{\omega } -1 \right)}{\frac{1}{\sqrt{1-{{\left( \cos \vartheta -\tilde{\omega } \right)}^{2}}}}d\vartheta },\ \ \ \! 0\le \tilde{\omega }\le 2 \\
\end{matrix} \right., \nonumber \\
& = \frac{2}{\mathrm{\pi}} \int_{0}^{\arccos \left( \left| {\tilde{\omega }} \right|-1 \right)}{\frac{1}{\sqrt{1 \!-\! {{\left( \cos \vartheta \!-\! \left| {\tilde{\omega }} \right| \right)}^{2}}}}d\vartheta }, \ \! \left| {\tilde{\omega }} \right| \le 2 \ \!.
\end{align}

In order to get some insights from the expression, we further rewrite (\ref{DistortionFunction_EquiAngle}) as
\begin{align} {\label{DistortionFunction_EquiAngle3}}
\mathcal{W} \left( {\tilde{\omega }} \right)=\left\{ \begin{matrix}
   \frac{2}{{\mathrm{\pi}} } \! \int_{0}^{\frac{{\mathrm{\pi}} }{2}} {\frac{1}{\sqrt{1-({1-\frac{{{{\tilde{\omega }}}^{2}}}{4}} ) {{\sin }^{2}}\xi }}d\xi }, \ 0\!<\! \left| {\tilde{\omega }} \right|\!<\!2  \\
   \infty, \hspace{10.85em} \ {\tilde{\omega }} \!=\! 0  \\
   1, \hspace{10.9em} \ \left| {\tilde{\omega }} \right| \!=\! 2 \\
\end{matrix} \right..
\end{align}
The detailed derivation can be found in Appendix A. As (\ref{DistortionFunction_EquiAngle3}) reveals, on the one hand, $\mathcal{W} \left( {\tilde{\omega }} \right)$ monotonically decreases with the increase of $|\tilde{\omega}|$; on the other hand, $\mathcal{W} \left( {\tilde{\omega }} \right)$ approaches to infinity and 1 as $|\tilde{\omega}|$ goes to 0 and 2, respectively. Thus, based on (\ref{DistortionFunction_EquiAngle3}), the profile of the beam-distortion function can be easily outlined.

\begin{remark}
For the generalized channel in \emph{Remark 1}, where the signal AoDs are distributed within $\left( {{\theta }_{\mathrm{L}}},{{\theta }_{\mathrm{R}}} \right)$, the beam-distortion functions in the above two cases can be similarly derived.
The results are summarized in Table~\ref{Table_DistortionFunction}, which compares different forms of beam-distortion functions under different channel assumptions and beamforming directions.
Especially, if uniform channel PAS is assumed for the generalized channel, the beam-distortion functions with `Equi-cos' and `Equi-angle' beamforming distrbutions reduce to
\begin{align} {\label{DistortionFunction_EquiCos2}}
\mathcal{W}\left( {\tilde{\omega }} \right) \!=\! \left\{ \begin{matrix}
  \! \frac{2\mathrm{\pi}} {{{\theta }_{\mathrm{R}}} - {{\theta }_{\mathrm{L}}}} \frac{\arccos \left( \cos {{\theta }_{\mathrm{R}}} -\tilde{\omega } \right) -{{\theta }_{\mathrm{L}}}} {\mu \left( {{\theta }_{\mathrm{L}}}, {{\theta }_{\mathrm{R}}} \right)}, \ -\mu \left( {{\theta }_{\mathrm{L}}}, {{\theta }_{\mathrm{R}}} \right) \!\le\! \tilde{\omega } \!<\! 0  \\
  \! \frac{2\mathrm{\pi}}{{{\theta }_{\mathrm{R}}} - {{\theta }_{\mathrm{L}}}} \frac{{{\theta }_{\mathrm{R}}}-\arccos \left( \cos {{\theta }_{\mathrm{L}}}-\tilde{\omega } \right)}{\mu \left( {{\theta }_{\mathrm{L}}},{{\theta }_{\mathrm{R}}} \right)},\ \ \ \ \! 0 \!\le\! \tilde{\omega } \!\le\! \mu \left( {{\theta }_{\mathrm{L}}},{{\theta }_{\mathrm{R}}} \right) \\
\end{matrix} \right.,
\end{align}
and
%
\begin{align} {\label{DistortionFunction_EquiAngle2}}
& \mathcal{W}\left( {\tilde{\omega }} \right) \!=\! \left\{ \begin{matrix}
 \!\! \frac{2\mathrm{\pi}}{({{\theta }_{\mathrm{R}}} - {{\theta }_{\mathrm{L}}})^{2}} \! \int_{\arccos \left(\tilde{\omega } + \cos {{\theta }_{\mathrm{L}}} \right)}^{{{\theta }_{\mathrm{R}}}}{\frac{1}{\sqrt{1-{{\left( \cos \vartheta -\tilde{\omega } \right)}^{2}}}}d\vartheta },\ \tilde{\omega }\!<\!0  \\
 \! \frac{2\mathrm{\pi}}{({{\theta }_{\mathrm{R}}} - {{\theta }_{\mathrm{L}}})^{2}} \! \int_{{{\theta }_{\mathrm{L}}}}^{\arccos \left(\tilde{\omega } + \cos {{\theta }_{\mathrm{R}}} \right)}{\frac{1}{\sqrt{1-{{\left( \cos \vartheta -\tilde{\omega } \right)}^{2}}}}d\vartheta },\ \! \tilde{\omega } \!\ge\! 0 \\
\end{matrix} \right.\!, \nonumber \\
& \hspace{40mm} \left| \tilde{\omega } \right| \le \mu \left( {{\theta }_{\mathrm{L}}},{{\theta }_{\mathrm{R}}} \right).
\end{align}
\end{remark}

\begin{table*}[!t]
\caption{ Beam-distortion functions under difference channel assumptions and beamforming directions }\label{Table_DistortionFunction}
\centering
\begin{tabular}{|c||c|c|}
\hline
$\mathcal{W}(\omega) $ &Jakes' channel & Non-Jakes' channel \\
\hline
Equi-cos & $\arccos \left( \left| {\tilde{\omega }} \right| \!-\! 1 \right), \ \! \left| {\tilde{\omega }} \right| \le 2$
& $\left\{ \begin{matrix}
   \frac{2\mathrm{\pi}} {\mu \left( {{\theta }_{\mathrm{L}}},{{\theta }_{\mathrm{R}}} \right)} \int_{{{\theta }_{\mathrm{L}}}}^{\arccos \left( \cos {{\theta }_{\mathrm{R}}}-\tilde{\omega } \right)}{ \rho(x) dx},\ -\mu \left( {{\theta }_{\mathrm{L}}},{{\theta }_{\mathrm{R}}} \right) \le \tilde{\omega }<0  \\
   \frac{2\mathrm{\pi}} {\mu \left( {{\theta }_{\mathrm{L}}},{{\theta }_{\mathrm{R}}} \right)} \int_{\arccos \left( \cos {{\theta }_{\mathrm{L}}}-\tilde{\omega } \right)}^{{{\theta }_{\mathrm{R}}}}{ \rho(x) dx},\ \ \ \ 0\le \tilde{\omega }\le \mu \left( {{\theta }_{\mathrm{L}}},{{\theta }_{\mathrm{R}}} \right) \\
\end{matrix} \right.$ \\
\hline
\!Equi-angle\!\! & $\left\{ \begin{matrix}
   \! \frac{2}{{\mathrm{\pi}} } \! \int_{0}^{\frac{{\mathrm{\pi}} }{2}} \! {\frac{1}{\sqrt{1-({1-\frac{{{{\tilde{\omega }}}^{2}}}{4}} ) {{\sin }^{2}}\xi }}d\xi }, 0\!<\! \left| {\tilde{\omega }} \right|\!<\!2  \\
   \infty, \! \hspace{11.3em} {\tilde{\omega }} \!=\! 0  \\
   1, \hspace{11.4em}\! \left| {\tilde{\omega }} \right| \!=\! 2  \\
\end{matrix} \right.$
& $\left\{ \begin{matrix}
   \! \frac{2\mathrm{\pi}}{{{\theta }_{\mathrm{R}}} - {{\theta }_{\mathrm{L}}}} \!\! \int_{\arccos \left(\tilde{\omega } + \cos {{\theta }_{\mathrm{L}}} \right)}^{{{\theta }_{\mathrm{R}}}} \! {\frac{ \rho\left( \arccos \left( \cos {{\vartheta }_{q}} - \tilde{\omega } \right) \right) }{\sqrt{1-{{\left( \cos \vartheta - \tilde{\omega } \right)}^{2}}}}d\vartheta }, -\mu \! \left( {{\theta }_{\mathrm{L}}},{{\theta }_{\mathrm{R}}} \right) \!\le\! \tilde{\omega }\!<\!0  \\
   \! \frac{2\mathrm{\pi}}{{{\theta }_{\mathrm{R}}} - {{\theta }_{\mathrm{L}}}} \!\! \int_{{{\theta }_{\mathrm{L}}}}^{\arccos \left(\tilde{\omega } + \cos {{\theta }_{\mathrm{R}}} \right)} \! {\frac{ \rho\left( \arccos \left( \cos {{\vartheta }_{q}} - \tilde{\omega } \right) \right) }{\sqrt{1-{{\left( \cos \vartheta - \tilde{\omega } \right)}^{2}}}}d\vartheta },\ \ 0 \!\le \! \tilde{\omega } \!\le\! \mu \! \left( {{\theta }_{\mathrm{L}}},{{\theta }_{\mathrm{R}}} \right) \\
\end{matrix} \right.$ \\
\hline
\end{tabular}
\end{table*}

The beam-distortion functions given in (\ref{DistortionFunction_EquiCos}), (\ref{DistortionFunction_EquiAngle}) and (\ref{DistortionFunction_EquiCos2}) are depicted in Fig. 2. Note that (\ref{DistortionFunction_EquiCos}) and (\ref{DistortionFunction_EquiCos2}) adopt ``Equi-cos'' while (\ref{DistortionFunction_EquiAngle}) adopts ``Equi-angle''. Besides, Jakes' channel is assumed for (\ref{DistortionFunction_EquiCos}) and (\ref{DistortionFunction_EquiAngle}), whereas we take ${{\theta }_{\mathrm{L}}} \!=\! 0^{\circ}, \theta_{\mathrm{R}} \!=\! 90^{\circ}$ for (\ref{DistortionFunction_EquiCos2}).
All the three beam-distortion functions are nonnegative and decrease with increasing $ |\tilde{\omega} |$. Hence, they all attain the maximum at $\tilde{\omega} \!=\! 0$.
Apart from this, the following observations can be made:

\begin{figure}[t]
\setlength{\abovecaptionskip}{-2mm}
\setlength{\belowcaptionskip}{-5mm}
\begin{center}
\includegraphics[width=80mm]{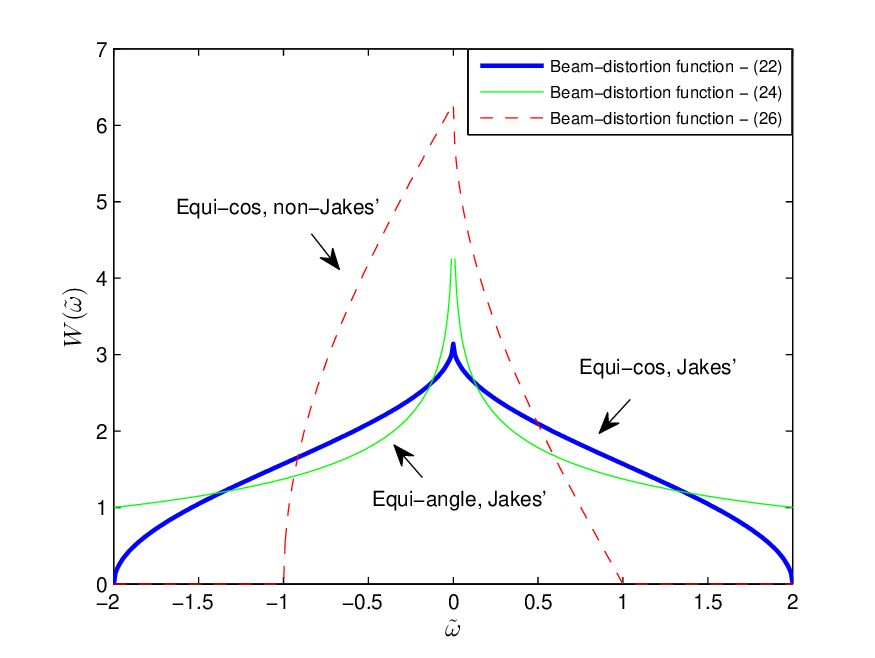}
\end{center}
\caption{ Comparison of the beam-distortion functions $W \left( {\tilde{\omega }} \right)$ given in (\ref{DistortionFunction_EquiCos}), (\ref{DistortionFunction_EquiAngle}) and (\ref{DistortionFunction_EquiCos2}) ($ {{\theta }_{\mathrm{L}}} \!=\! 0^{\circ}, \theta_{\mathrm{R}} \!=\! 90^{\circ}$ for (\ref{DistortionFunction_EquiCos2})). }
\end{figure}

First, unlike (\ref{DistortionFunction_EquiCos}) and (\ref{DistortionFunction_EquiAngle}), (\ref{DistortionFunction_EquiCos2}) yields a beam-distortion function which is asymmetric about $\tilde{\omega} \!=\! 0$.
Such asymmetry is due to the fact that the mean AoD deviates from $\frac{\mathrm{\pi}}{2}$.
Moreover, $\mathcal{W} \left( {\tilde{\omega }} \right)$ in (\ref{DistortionFunction_EquiCos2}) remains zero for $|\tilde{\omega}| \!>\! 1$, due to $\mu \left( {{\theta }_{\mathrm{L}}},{{\theta }_{\mathrm{R}}} \right) \!=\! 1$.
Second, comparing (\ref{DistortionFunction_EquiCos}) and (\ref{DistortionFunction_EquiAngle}), we observe that the beam-distortion function in (\ref{DistortionFunction_EquiAngle}) is more concentrated around $\tilde{\omega} \!=\! 0$, while that in (\ref{DistortionFunction_EquiCos}) better attenuates the high Doppler frequencies as $|\tilde{\omega}|$ approaches 2.
Third, $\mathcal{W} \left( {\tilde{\omega }} \right)$ in (\ref{DistortionFunction_EquiAngle}) is unbounded above at $\tilde{\omega} \!=\! 0$ and converges to 1 as $ |\tilde{\omega} |$ tends to 2, which matches with the analysis obtained from the alternative form (\ref{DistortionFunction_EquiAngle3}).

\subsubsection{Impact of antenna spacing}
As derived in \emph{Lemma 1}, the pattern function $| \mathcal{G} \left( {\tilde{\omega }} \right)\! |^{2}$ is given by $| \mathcal{G} \left( {\tilde{\omega }} \right) \!|^{2} \!=\! \frac{{{\sin }^{2}}\left( \chi M \tilde{\omega } \right)}{M^2 {{\sin }^{2}}\left( \chi \tilde{\omega } \right)}$. Apparently, $| \mathcal{G} \left( {\tilde{\omega }} \right)\! |^{2}$ manifests itself as a periodic function of $\tilde{\omega }$, which repeats itself with period $\tilde {\Omega} \!=\! \frac{\mathrm{\pi}}{\chi}$.

Note that the antenna spacing $d$ can be set a bit larger to gain higher beamforming resolution, but it cannot exceed $d_{\mathrm{max}} \!=\! \frac{\lambda}{2}$ to avoid aliasing. Moreover, $d \!=\! \frac{\lambda}{2}$ will also incur the aliasing between $0^{\circ}$ and $180^{\circ}$. Therefore, we limit the range of antenna spacing as $ 0 \!<\! d \!<\! \frac{\lambda}{2}$, and the optimal antenna spacing should be compromised between beamforming resolution and aliasing avoidance.

\setlength{\abovecaptionskip}{-0mm}
\setlength{\belowcaptionskip}{-5mm}
\begin{figure}[htbp]
\centering
\makeatletter\def\@captype{figure}\makeatother
  \subfigure{
    \includegraphics[width=80mm]{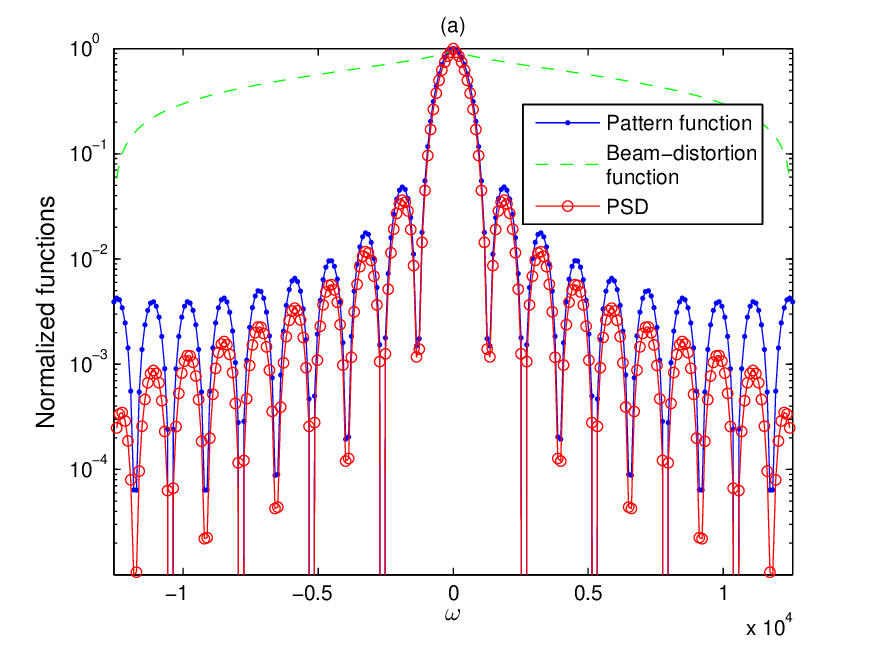}}
  \hspace{-0.2 in}
  \subfigure{
    \includegraphics[width=80mm]{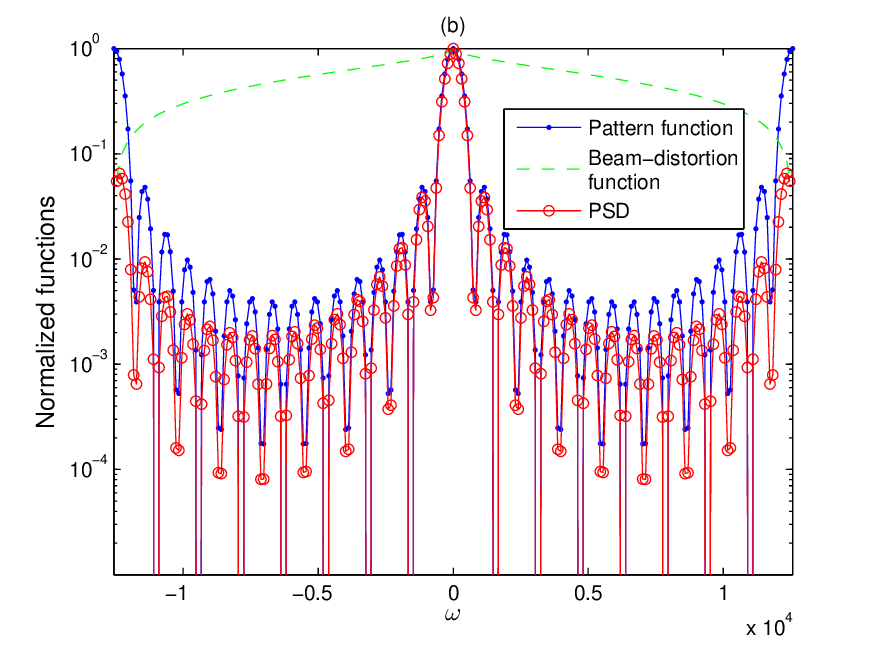}}
\caption{Comparison of the normalized pattern function $|\mathcal{G}(\omega)|^{2}$, beam-distortion function $W(\omega)$ and PSD $P(\omega)$, for 16-element ($M\!=\!16$) ULA with the normalized antenna spacings set as: (a) $\frac{d}{\lambda} = 0.3$ and (b) $\frac{d}{\lambda} = 0.5$. }
\end{figure}

Fig. 3 compares the pattern function $|\mathcal{G}(\omega)|^{2}$, beam-distortion function $\mathcal{W}(\omega)$ and PSD $P(\omega)$, when the antenna spacings are taken as $d \!=\! 0.3 \lambda$ and $d \!=\! 0.5 \lambda$, respectively. Note that the absolute values of $|\mathcal{G}(\omega)|^{2}$, $\mathcal{W}(\omega)$ and $P(\omega)$ have been scaled such that their maximums are all 1 (e.g., the depicted beam-distortion function is in fact $\mathcal{W}(\omega) / \max\{ |\mathcal{W}(\omega)| \}$).
The maximum Doppler shift is taken as $f_d \!=\! 1 \!\ 000 \ \mathrm{Hz}$. The beamforming directions are configured such that $\cos \vartheta_q, \!\ q\!=\! 1,2,\ldots, Q$, are uniformly distributed between $\left(-1,1 \right)$. Since the beamforming directions are exactly the same, both cases share the same beam-distortion function $\mathcal{W}(\omega)$. Hence, only the choice of antenna spacing accounts for the difference between Fig. 3(a) and Fig. 3(b).
As anticipated, the pattern function $|\mathcal{G}(\omega)|^{2}$ at $\frac{d}{\lambda} \!=\! 0.5$ accomplishes a full period within $\omega \!\in\! (-2\omega_d, 2\omega_d)$, which implies $|\mathcal{G}(\pm 2\omega_d)|^{2} \!=\! |\mathcal{G}(0)|^{2}$. Therefore, despite the attenuation effect of the beam-distortion function $W(\omega)$ on large Doppler frquencies, the PSD at $\frac{d}{\lambda} \!=\! 0.5$ would be much larger than that at $\frac{d}{\lambda} \!=\! 0.3$ for large $\omega$. Since a PSD concentrated around low Doppler frequencies is more favorable for reducing the residual channel time variation, the antenna spacing $d \!=\! 0.5\lambda$ should be avoided. This graphically explains from another perspective why the tradeoff between beamforming resolution and aliasing avoidance needs to be taken when determining the antenna spacing $d$.

\section{Beamforming Network Optimization for Reducing Doppler Spread}
In Section III, we have employed $\mathbb{B} \!=\! \big[ \mathbf{b}\left( {{\vartheta }_{1}} \right), \!\ \mathbf{b}\left( {{\vartheta }_{2}} \right), $ $\ldots,\!\  \mathbf{b}\left( {{\vartheta }_{Q}} \right) \big]$ as the beamforming network, with $\mathbf{b}\left( {{\vartheta }_{q}} \right) \!=\! \frac{\eta}{M\sqrt{Q}}{{\mathbf{a}}}\left( {{\vartheta }_{q}} \right){{\mathrm{e}}^{\mathrm{j}\phi \left( {{\vartheta }_{q}} \right)}}$, to separate the Doppler shifts and reduce channel time variation. However, the MF beamformers are amplitude-constrained and phase-quantized vectors, which are suboptimal themselves in suppressing the residual Doppler shifts.
If the Doppler shifts separation could be performed by the optimal beamformers with entirely configurable amplitudes and phases, the residual Doppler shifts could be minimized, further reducing the channel time variation. Yet, acquiring such optimal beamformers leads to the joint optimization of $Q(M\!-\!1)$ parameters, which is of prohibitively high computational complexity.
In fact, we could turn to optimize the MF beamforming network $\mathbb{B}$ by introducing a Common Configurable Amplitudes and Phases (CCAP) parameter\footnote{The designation of CCAP parameter for $\mathbf{u}$ is due to the fact that all the MF beamformers $\mathbf{b}\left( {{\vartheta }_{q}} \right)$'s share the same CCAP parameter, which could remove in some degree the constant modulus and quantized phase constraints $\mathbf{b}\left( {{\vartheta }_{q}} \right)$'s are subject to.}
$\mathbf{u} \!=\! {{\left[ \begin{matrix}
   {{u}_{1}}, & {{u}_{2}}, & \ldots, & {{u}_{M}}  \\
\end{matrix} \right]}^{T}}$ such that
\begin{align}
\mathbb{B}_{\mathrm{CCAP}} &= \left[ \mathbf{b}_{\mathrm{CCAP}}\left( {{\vartheta }_{1}} \right),\ \mathbf{b}_{\mathrm{CCAP}}\left( {{\vartheta }_{2}} \right),\ \ldots, \ \mathbf{b}_{\mathrm{CCAP}}\left( {{\vartheta }_{Q}} \right) \right] \nonumber \\
& = \operatorname{diag} \left( \mathbf{u}^{*} \right) \mathbb{B}.
\end{align}
The $q$th beamformer is thus given by $\mathbf{b}_{\mathrm{CCAP}} \left( {{\vartheta }_{q}} \right) \!=\! \frac{\eta_{\mathrm{CCAP}}}{M\sqrt{Q}} \operatorname{diag} \left( {\mathbf{u}}^{*} \right) {{\mathbf{a}}} \left( {{\vartheta }_{q}} \right) {{\mathrm{e}}^{\mathrm{j}\phi \left( {{\vartheta }_{q}} \right)}}$, where $\eta_{\mathrm{CCAP}} \!=\! \frac{1}{\left\| \sum\limits_{q=1}^{Q} {\frac{1}{M\sqrt{Q}} \operatorname{diag} \left( \mathbf{u}^{*} \right) \mathbf{a}\left( {{\vartheta }_{q}} \right) {{\mathrm{e}}^{\mathrm{j}\phi \left( {{\vartheta }_{q}} \right)}}} \right\|_2}$ is the normalization coefficient to keep the total transmit power per symbol as 1.

Note that the channel time variation can be reflected by Doppler spread, the smaller the Doppler spread is, the slower the channel varies in time. As a result, we could obtain the optimal CCAP parameter by minimizing the Doppler spread. To this end, we must first derive the channel PSD with the modified beamformers $\mathbf{b}_{\mathrm{CCAP}} \left( {{\vartheta }_{q}} \right)$, and it is expected that the CCAP parameter $\mathbf{u}$ will only affect the pattern function, since the pattern function corresponds to the converted radiation pattern of the array.

\begin{figure}[t]
\setlength{\abovecaptionskip}{-0mm}
\setlength{\belowcaptionskip}{-5mm}
\begin{center}
\includegraphics[width=90mm]{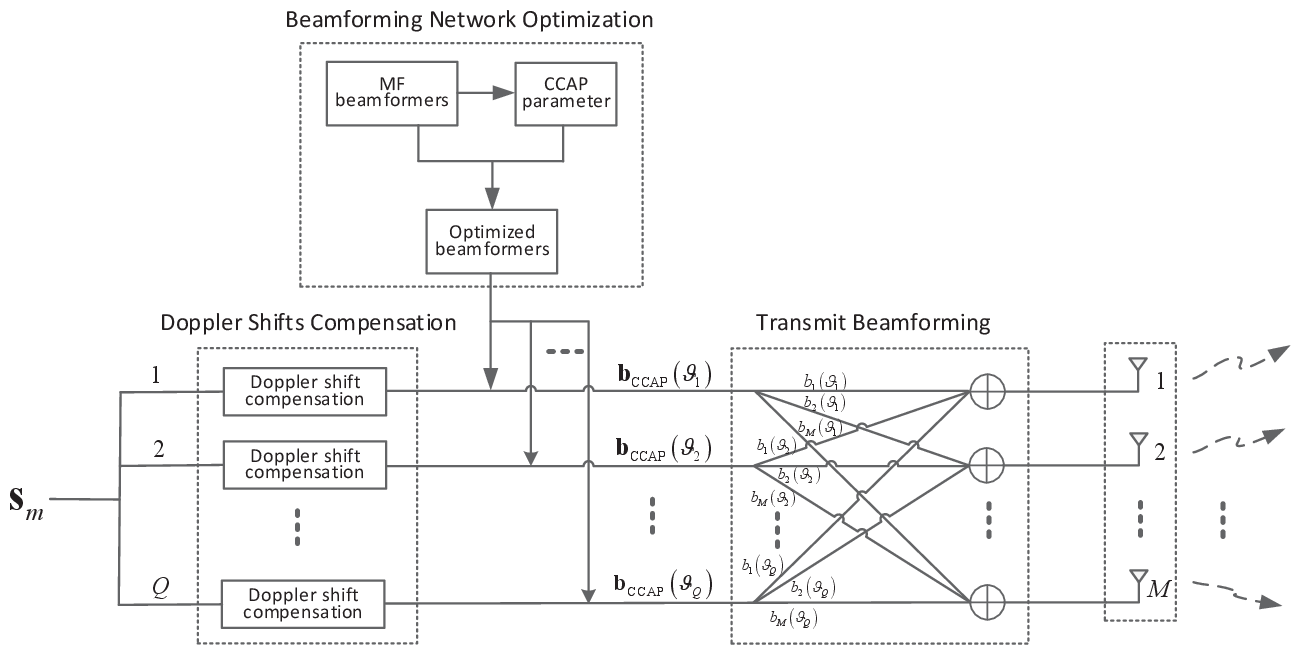}
\end{center}
\caption{ Illustration of the beamforming network optimization technique. }
\end{figure}

With the new beamforming network $\mathbb{B}_{\mathrm{CCAP}}$, the signal received at the BS after Doppler shifts compensation and multi-branch transmit beamforming can be re-expressed as (the noise item is ignored and only one channel tap is considered)
\begin{align}{\label{ReceivedBeamSignal2}}
{{\mathbf{r}}_{m, {\mathrm{CCAP}}}} & =\sum\limits_{q=1}^{Q}\sum\limits_{p=1}^{P} {{\rho }_{1,p}} \mathbf{a}^{T}\left( {{\theta }_{1,p}} \right) {{\mathbf{b}}^{*}_{\mathrm{CCAP}}} \left( {{\vartheta }_{q}} \right) {{\mathbf{s}}_{m}} \left( {{d}_{1}} \right) \nonumber \\
& \hspace{20mm} \times {{\mathbf{\Psi }}_{m,1}} \left( {{\vartheta }_{q}} \right) {{\mathbf{\Phi }}_{m}} \left( \theta_{1,p} \right), \nonumber \\
& =\frac{\eta_{\mathrm{CCAP}} }{\sqrt{Q}}\sum\limits_{q=1}^{Q} \sum\limits_{p=1}^{P}{{\rho }_{p}} {{\mathrm{e}}^{-\mathrm{j}\phi \left( {{\vartheta }_{q}} \right)}} \frac{1}{M} \mathbf{a}^{H} \left( {{\vartheta }_{q}} \right) \operatorname{diag} \left( \mathbf{u} \right) \nonumber \\
& \hspace{20mm} \times {{\mathbf{a}}}\left( {{\theta }_{p}} \right) {{\mathbf{s}}_{m}} {{\mathbf{\Psi }}_{m,1}} \left( {{\vartheta }_{q}} \right) {{\mathbf{\Phi }}_{m}}\left( \theta_p \right).
\end{align}

By ignoring the real scalar $\eta_{\mathrm{CCAP}} $ which does not affect the PSD analysis and Doppler spread, the equivalent uplink channel of (\ref{ReceivedBeamSignal2}) can be expressed in continuous-time form as
\begin{align}{\label{ContinuousChannel2}}
& g_{\mathrm{CCAP}} \left( t \right) =\frac{1}{\sqrt{Q}}\sum\limits_{q=1}^{Q} \int_{0}^{\mathrm{2\pi}} \alpha \left( \theta  \right) G_{\mathrm{CCAP}}\left( \cos \theta ,\cos{{\vartheta }_{q}} \right) \nonumber \\
& \hspace{18mm} \times {{\mathrm{e}}^{\mathrm{j}{{\omega }_{d}}\left( \cos \theta -\cos {{\vartheta }_{q}} \right)t+\mathrm{j}\varphi \left( \theta  \right)-\mathrm{j}\phi \left( {{\vartheta }_{q}} \right)}}d\theta,
\end{align}
where
\begin{align} {\label{G_CCAP}}
& G_{\mathrm{CCAP}} \left( \cos \theta ,\cos{{\vartheta }_{q}} \right) = \frac{1}{M} \mathbf{a}^{H}\left( {{\vartheta }_{q}} \right) \operatorname{diag} \left( \mathbf{u} \right) \mathbf{a}\left( \theta  \right) \nonumber \\
& \hspace{20mm} = \frac{1}{M} \sum\nolimits_{r=1}^{M}{u_r {{\mathrm{e}}^{\mathrm{j2}\chi (r-1) \left( \cos \theta -\cos {{\vartheta }_{q}} \right)}}}.
\end{align}
By denoting $\mathbf{c}\left( \cos \theta ,\cos {{\vartheta }_{q}} \right) \!=\! \big[ 1,\ {{\mathrm{e}}^{\mathrm{j2} \chi \left( \cos \theta -\cos {{\vartheta }_{q}} \right)}},\ \ldots,$ $ {{\mathrm{e}}^{\mathrm{j2} \chi (M-1) \left( \cos \theta -\cos {{\vartheta }_{q}} \right)}} \big]^{T}$, (\ref{G_CCAP}) could be rewritten as
\begin{align}
G_{\mathrm{CCAP}} \left( \cos \theta ,\cos{{\vartheta }_{q}} \right) = \frac{1}{M}{{\mathbf{c}}^{T}}\left( \cos \theta ,\cos {{\vartheta }_{q}} \right)\mathbf{u}.
\end{align}

Note that the only difference between the continuous-time channels in (\ref{ContinuousChannel2}) and (\ref{ContinuousChannel}) is that $G \left( \cos \theta ,\cos{{\vartheta }_{q}} \right)$ in (\ref{ContinuousChannel}) is replaced by $G_{\mathrm{CCAP}} \left( \cos \theta ,\cos{{\vartheta }_{q}} \right)$ in (\ref{G_CCAP}). Actually, by letting $\mathbf{u} \!=\! {\mathbf{1}}_{M\times1}$, $G_{\mathrm{CCAP}} \left( \cos \theta ,\cos{{\vartheta }_{q}} \right)$ will reduce to $G\left( \cos \theta ,\cos{{\vartheta }_{q}} \right)$. Thus, the considered scenario in Section III can be categoried as a special case of directly adopting MF beamformers without configurable amplitudes and phases (or with all-one CCAP parameter). Following the similar approach as in Section III, the channel PSD can be expressed as
\begin{align}{\label{PSD_CCAP}}
P_{\mathrm{CCAP}} \left( \omega  \right) = \frac{1}{{{\omega }_{d}}}  {{\left| {\mathcal{G}}_{\mathrm{CCAP}} (\tilde{\omega}) \right|}^{2}} \mathcal{W}\left( {\tilde{\omega }} \right),
\end{align}
where the beam-distortion function remains exactly the same as (\ref{DistortionFunction}), implying that the nonzero PSD region is still $\left| \tilde{\omega} \right| \le 2$, while the pattern function can be redefined as
\begin{align}{\label{PatternFunction_CCAP}}
| {\mathcal{G}}_{\mathrm{CCAP}} (\tilde{\omega})|^{2} & = | \frac{1}{M} \sum\nolimits_{r=1}^{M} { u_r {{\mathrm{e}}^{-\mathrm{j2} \chi (r-1) \tilde{\omega }}}} |^{2} \nonumber \\
&  = | \frac{1}{M}{{\boldsymbol{\varsigma }}^{H}}\left(  \tilde{\omega } \right)\mathbf{u} |^{2}.
\end{align}
Here, $ \boldsymbol{\varsigma } \left( \tilde{\omega } \right) \!=\! {{\left[ 1,\ {{\mathrm{e}}^{\mathrm{j2} \chi \tilde{\omega } }},\ \ldots ,\ {{\mathrm{e}}^{\mathrm{j2} \chi (M-1) \tilde{\omega } }} \right]}^{T}}$.

As a result, the Doppler spread with the beamforming network $\mathbb{B}_{\mathrm{CCAP}}$ can be calculated as
\begin{align} {\label{DopplerSpread_CCAP}}
{{\sigma }_{\mathrm{DS, CCAP}}} & =\sqrt{\frac{\int_{-2\ \!\! {{\omega }_{d}}}^{2\ \!\! {{\omega }_{d}}}{{{\omega }^{2}}P_{\mathrm{CCAP}} \left( \omega  \right)d\omega }}{\int_{-2\ \!\! {{\omega }_{d}}}^{2\ \!\!  {{\omega }_{d}}}{P_{\mathrm{CCAP}} \left( \omega  \right)d\omega }}}, \nonumber \\
& \hspace{-2mm} \overset{\omega ={{\omega }_{d}}\tilde{\omega }}{\mathop{=}}\, \  {{\omega }_{d}}\sqrt{\frac{\int_{-2}^{2 }{{{{\tilde{\omega }}}^{2}}{{\left| {{\boldsymbol{\varsigma }}^{H}}\left(  \tilde{\omega } \right)\mathbf{u} \right|}^{2}} \mathcal{W}\left( {\tilde{\omega }} \right)d\tilde{\omega }}}{\int_{-2}^{2} {{{\left| {{\boldsymbol{\varsigma }}^{H}}\left(  \tilde{\omega } \right)\mathbf{u} \right|}^{2}} \mathcal{W}\left( {\tilde{\omega }} \right)d\tilde{\omega }}}}, \nonumber \\
& = {{\omega }_{d}} \sqrt{\frac{{{\mathbf{u}}^{H}} \left[ \int_{-2}^{2} {{{{\tilde{\omega }}}^{2}} {\mathcal{W}} \left( {\tilde{\omega }} \right){{\boldsymbol{\varsigma }}}\left( {\tilde{\omega }} \right) {{\boldsymbol{\varsigma }}^{H}} \left( {\tilde{\omega }} \right) d\tilde{\omega }} \right] \mathbf{u}}{{{\mathbf{u}}^{H}}\left[ \int_{-2}^{2}{{\mathcal{W}} \left( {\tilde{\omega }} \right) {{\boldsymbol{\varsigma }}}\left( {\tilde{\omega }} \right) {{\boldsymbol{\varsigma }}^{H}} \left( {\tilde{\omega }} \right)d\tilde{\omega }} \right]\mathbf{u}}}, \nonumber \\
& = {{\omega }_{d}} \sqrt{\frac{{{\mathbf{u}}^{H}} {{\mathbf{C}}_{2}} \mathbf{u}} {{{\mathbf{u}}^{H}} {{\mathbf{C}}_{0}} \mathbf{u}}},
\end{align}
where
\begin{align*}
\mathbf{C}_0 & = \int_{-2}^{2}{ {\mathcal{W}} \left( {\tilde{\omega }} \right){{\boldsymbol{\varsigma }}}\left( {\tilde{\omega }} \right){{\boldsymbol{\varsigma }}^{H}}\left( {\tilde{\omega }} \right)d\tilde{\omega }}, \nonumber \\
\mathbf{C}_2 & = \int_{-2}^{2}{{{{\tilde{\omega }}}^{2}} {\mathcal{W}} \left( {\tilde{\omega }} \right){{\boldsymbol{\varsigma }}}\left( {\tilde{\omega }} \right){{\boldsymbol{\varsigma }}^{H}}\left( {\tilde{\omega }} \right)d\tilde{\omega }}.
\end{align*}
Note that when the beam-distortion function ${\mathcal{W}} \left( {\tilde{\omega }} \right)$ is symmetric about ${\tilde{\omega }} = 0$, both $\mathbf{C}_0$ and $\mathbf{C}_2$ are real symmetric Toeplitz matrices.

The optimal CCAP parameter $\hat{\mathbf{u}}$ minimizing the Doppler spread can be acquired by solving the following optimization problem
\begin{align} {\label{Opt_Problem}}
\hat{\mathbf{u}} = & \arg \underset{{\tilde{\mathbf{u}}}}{\mathop{\min }}\,\sigma _{\mathrm{DS,CCAP}}
=\arg \underset{{\tilde{\mathbf{u}}}}{\mathop{\min }}\, \frac{{{{\tilde{\mathbf{u}}}}^{H}} {{\mathbf{C}}_{2}} {\tilde{\mathbf{u}}}} {{{{\tilde{\mathbf{u}}}}^{H}}{{\mathbf{C}}_{0}} {\tilde{\mathbf{u}}}},\nonumber \\
& \!\ s.t. \!\ {{{\tilde{\mathbf{u}}}}^{H}} {{\mathbf{C}}_{0}} {\tilde{\mathbf{u}}}=1,
\end{align}
where ${\tilde{\mathbf{u}}}$ denotes the trial CCAP parameter, and the constraint ${{{\tilde{\mathbf{u}}}}^{H}} {{\mathbf{C}}_{0}} {\tilde{\mathbf{u}}} \!=\!1$ is added to eliminate the magnitude ambiguity of $\tilde{\mathbf{u}}$ and also to avoid the trivial solution of $\hat{\mathbf{u}} \!=\! \mathbf{0}$.

The optimization problem in (\ref{Opt_Problem}) is a typical Rayleigh-entropy problem~\cite{Zhang2017}, and the optimal CCAP parameter minimizing the Doppler spread can be obtained in a closed form as
\begin{align}{\label{SolutionCCAP}}
\hat{\mathbf{u}}= {{\mathbf{Q}}^{-H}} {{\mathbf{v}}_{\min }} \left( {{\mathbf{Q}}^{-1}} {{\mathbf{C}}_{2}} {{\mathbf{Q}}^{-H}} \right),
\end{align}
where $\mathbf{Q}$ is acquired by decomposing ${{\mathbf{C}}_{0}}$ as ${{\mathbf{C}}_{0}} \!=\! \mathbf{Q}{{\mathbf{Q}}^{H}}$ and ${{\mathbf{v}}_{\min }} \left( \mathbf{X} \right)$ denotes the eigenvector corresponding to the minimum eigenvalue of matrix $\mathbf{X}$. The $q$th optimal beamformer can finally be expressed as $\hat{\mathbf{b}}_{\mathrm{CCAP}} \left( {{\vartheta }_{q}} \right) \!=\! \frac{\eta_{\mathrm{CCAP}}}{M\sqrt{Q}} \operatorname{diag} \left( \hat{\mathbf{u}}^{*} \right) {{\mathbf{a}}} \left( {{\vartheta }_{q}} \right) {{\mathrm{e}}^{\mathrm{j}\phi \left( {{\vartheta }_{q}} \right)}}$.

\begin{remark}
It can be seen from (\ref{Opt_Problem}) that the optimization of CCAP parameter does not depend on the maximum Doppler shift $f_d$ and thus is independent of the velocity of the HSR. Besides, the optimal CCAP parameter $\hat{\mathbf{u}}$ in (\ref{SolutionCCAP}) can be expressed in a closed form as a function of $\mathbf{C}_0$ and $\mathbf{C}_2$. Thus, for a given ULA with fixed antenna spacing, the optimal CCAP parameter $\hat{\mathbf{u}}$ is uniquely determined by the beam-distortion function $\mathcal{W}\left( \tilde{\omega} \right)$, which depends on the beamforming directions. In other words, as long as the beamforming network is determined, the CCAP parameter can be optimized and the obtained $\hat{\mathbf{u}}$ remains valid irrespective of HSR velocity.
\end{remark}

\begin{figure}[t]
\setlength{\abovecaptionskip}{-1mm}
\setlength{\belowcaptionskip}{-5mm}
\begin{center}
\includegraphics[width=80mm]{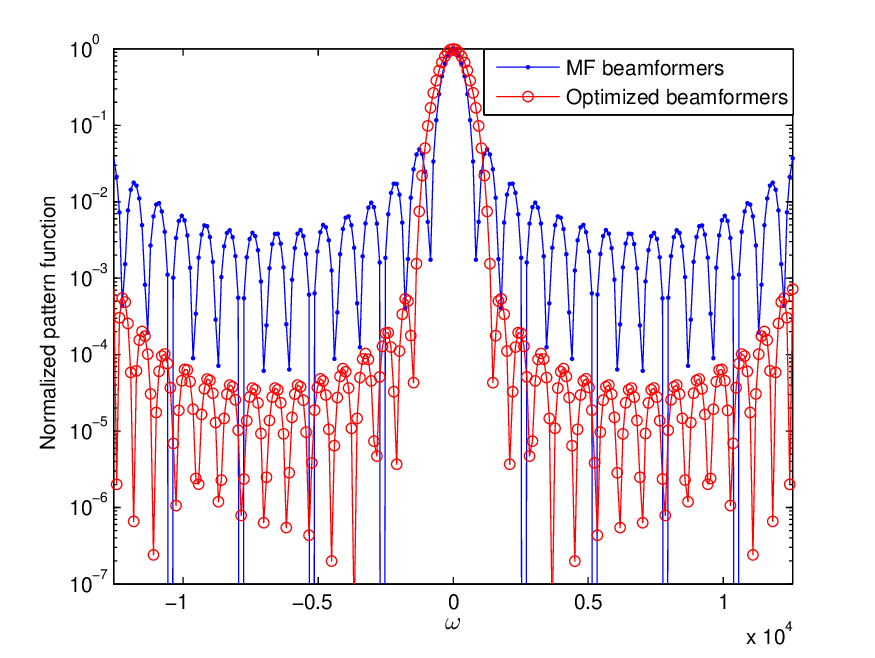}
\end{center}
\caption{ Comparison of the normalized pattern functions $|{\mathcal{G}}(\omega)|^{2}$ (MF beamformers) and $|{\mathcal{G}}_{\mathrm{CCAP}}(\omega)|^{2}$ (optimized CCAP beamformers) for $M\!=\!16$-element ULA, with antenna spacing $d=0.45\lambda$ and ``Equi-cos'' beamforming directions. }
\end{figure}

As expected, the introduction of the CCAP parameter will affect the pattern function. Hence, we compare in Fig. 5 the pattern function $|{\mathcal{G}}_{\mathrm{CCAP}}(\omega)|^{2}$ obtained with the beamformers ${\hat{\mathbf{b}}_{\mathrm{CCAP}}}\left( {{\vartheta }_{q}} \right)$ optimized by $\hat{\mathbf{u}}$ and $|\mathcal{G}(\omega)|^{2}$ obtained with MF beamformers ${{\mathbf{b}}}\left( {{\vartheta }_{q}} \right)$.
The 16-element ULA with antenna spacing $d \!=\! 0.45 \lambda$ is adopted, and the maximum Doppler shift is $f_d \!=\! 1 \!\ 000 \ \mathrm{Hz}$. Note that the absolute values of pattern functions are scaled such that their maximums are all 1, as in Fig. 3.
For each pattern function, we define the ratio between the sidelobe levels and the maximum gain as side-to-main ratio (SMR) $\rho$. It can be seen that the average SMR $\rho$ is about $10^{-2}$ for $|\mathcal{G}(\omega)|^{2}$ with MF beamformers, while the pattern function $|{\mathcal{G}}_{\mathrm{CCAP}}(\omega)|^{2}$ with the beamformers optimized by $\hat{\mathbf{u}}$ yields an average SMR of $\rho_{\mathrm{CCAP}} \approx 10^{-4}$, two orders of magnitude smaller than the former. Such low SMR is obtained at the cost of a slightly wider mainlobe.
Nevertheless, the SMR has greater impact on Doppler spread than the mainlobe width. Hence, the Doppler spread could be significantly reduced with the optimal CCAP beamformers, which substantially attenuate the high Doppler frequencies.

\begin{remark}
The result in Fig. 5 reveals that compared to the pure MF beamformers ${{\mathbf{b}}}\left( {{\vartheta }_{q}} \right)$, the optimal beamformers ${\hat{\mathbf{b}}_{\mathrm{CCAP}}}\left( {{\vartheta }_{q}} \right)$ incorporating the CCAP parameter $\hat{\mathbf{u}}$ can better eliminate the residual Doppler shifts and reduce the channel time variation. This is achieved via equivalently modifying the radiation pattern.
However, our proposed beamforming network optimization technique is different from the traditional array pattern synthesis\footnote{In general, array pattern synthesis refers to achieve the desired array radiation pattern with explicit specifications by designing the amplitude and phase excitation of different antennas~\cite{Lebret_TSP1997, Shi_SPL2005, Wang_TWC2016}. In~\cite{Lebret_TSP1997} for example, the amplitude and phase excitations are optimized to minimize the beam pattern level over a given region under other constraints.}.
Unlike array pattern synthesis, we do not have a priori a desired radiation pattern to attain, neither can we determine an ``optimal'' radiation pattern as a criterion for designing the CCAP parameter. Instead, the CCAP parameter is introduced to remove the constraints on MF beamformers to some degree and further optimized by minimizing the Doppler spread, resulting in the modified array radiation pattern. 
\end{remark}

\begin{remark}
The proposed beamforming network optimization technique can be equally applied to the generalized channel in \emph{Remark 1}, where the signal AoDs are distributed within $\left( {{\theta }_{\mathrm{L}}},{{\theta }_{\mathrm{R}}} \right)$.
Note that when the realistic AoD range $\left( {{\theta }_{\mathrm{L}}},{{\theta }_{\mathrm{R}}} \right)$ is not perfectly known, we can perform beamforming towards a slightly wider AoD range $\left( \tilde{\theta}_{\mathrm{L}}, \tilde{\theta}_{\mathrm{R}} \right)$ with $\tilde{\theta}_{\mathrm{L}} \!<\! \theta_{\mathrm{L}}, \tilde{\theta}_{\mathrm{R}} \!>\! \theta_{\mathrm{R}}$ to cover the realistic range.
In particular, in the worst case where the AoD range is completely unknown, we can simply employ the optimized beamforming network obtained under Jakes' channel to perform beamforming towards $\left( 0, \mathrm{\pi} \right)$.
\end{remark}

\section{Simulation Results}
In this section, we first verify the accuracy of channel PSD analysis and investigate the impact of some critical parameters on channel PSD and Doppler spread through numerical examples, and then demonstrate the superiority of the proposed beamforming network optimization technique over the simplest MF beamfoming network. Unless otherwise stated, the antenna spacing is taken as $d\!=\!0.45{\lambda}$, the maximum Doppler shift is set as $f_d \!=\! 1\!\ 000\ \mathrm{Hz}$ and the ULA consists of $M\!=\!16$ antennas.

\subsection{Verification of Channel PSD Analysis}

\setlength{\abovecaptionskip}{-1mm}
\setlength{\belowcaptionskip}{-3mm}
\begin{figure}[htbp]
\centering
\makeatletter\def\@captype{figure}\makeatother
  \subfigure{
    \includegraphics[width=80mm]{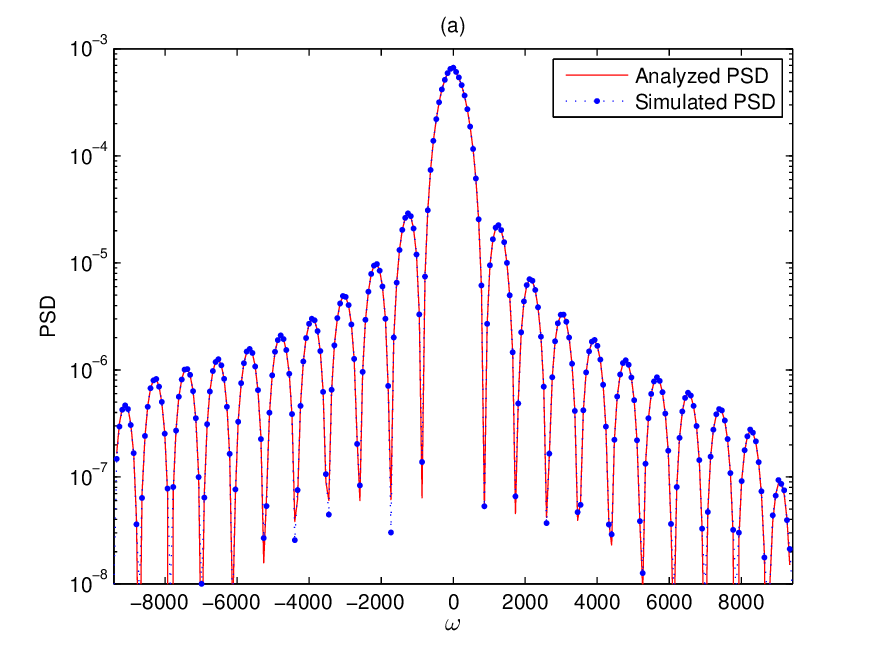}}
  \hspace{-0.2 in}
  \subfigure{
    \includegraphics[width=80mm]{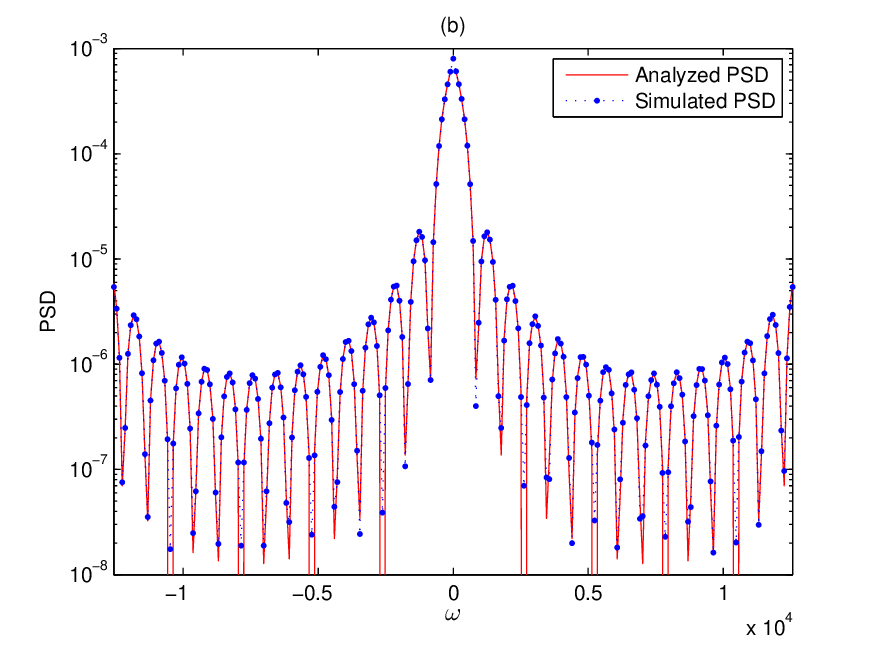}}
\caption{ Comparison of the PSD $P(\omega)$ between two circumstances of (a) Non-Jakes' channel ($\theta_{\mathrm{L}} \!=\! 0^{\circ}, \ \theta_{\mathrm{R}} \!=\! 120^{\circ}$) with ``Equi-cos'' beamforming directions and (b) Jakes' channel with ``Equi-angle'' beamforming directions. }
\end{figure}

In Fig. 6, we compare the channel PSD under different channel assumptions and beamforming directions. The AoDs are constrained within $({{\theta }_{\mathrm{L}}}, {{\theta }_{\mathrm{R}}} )$ with ${{\theta }_{\mathrm{L}}} \!=\! 0^{\circ}, \theta_{\mathrm{R}} \!=\! 120^{\circ}$ and the beamforming directions $\vartheta_q$ are configured such that $\cos \vartheta_q$ are evenly distributed between $\left( \cos {{\theta }_{\mathrm{R}}}, \cos {{\theta }_{\mathrm{L}}} \right)$ in Fig. 6(a), while Jakes' channel model and `Equi-angle' beamforming directions are adopted for Fig. 6(b). That is to say, (\ref{DistortionFunction_EquiCos2}) and (\ref{DistortionFunction_EquiAngle}) should be employed to compute the beam-distortion function, respectively.

In order to verify the correctness of the PSD derivation (\ref{PSD}), we provide the numerical PSD obtained in the following way:
We first calculate the channel autocorrelation ${{R}_{g}}\left( \tau  \right)$ at $T$ discrete time points by averaging over sufficient number of channel realizations (\ref{ContinuousChannel}) and then apply a $T$-point discrete Fourier transform (DFT) to obtain the discretized PSD.
Note that $T$ should be accordingly increased with the number of antennas $M$ to capture the faster fluctuation of the magnitude of PSD.
Fig. 6 reveals that whether the channel follows Jakes' model or not, the analyzed PSD (\ref{PSD}) perfectly coincides with its numerical counterpart, confirming the validity of the PSD analysis.
Furthermore, we can find that the PSD in Fig. 6(a) is asymmetric about $\omega\!=\!0$ while that in Fig. 6(b) is symmetric. In fact, we have pointed out in Fig. 2 that an average AoD different from $\frac{\mathrm{\pi}}{2}$ will result in asymmetric beam-distortion function, which accounts for the asymmetry of the PSD in Fig. 6(a).

\subsection{Influence of Some Critical Parameters on Channel PSD and Doppler Spread}
\begin{figure}[t]
\setlength{\abovecaptionskip}{-1mm}
\setlength{\belowcaptionskip}{-5mm}
\begin{center}
\includegraphics[width=80mm]{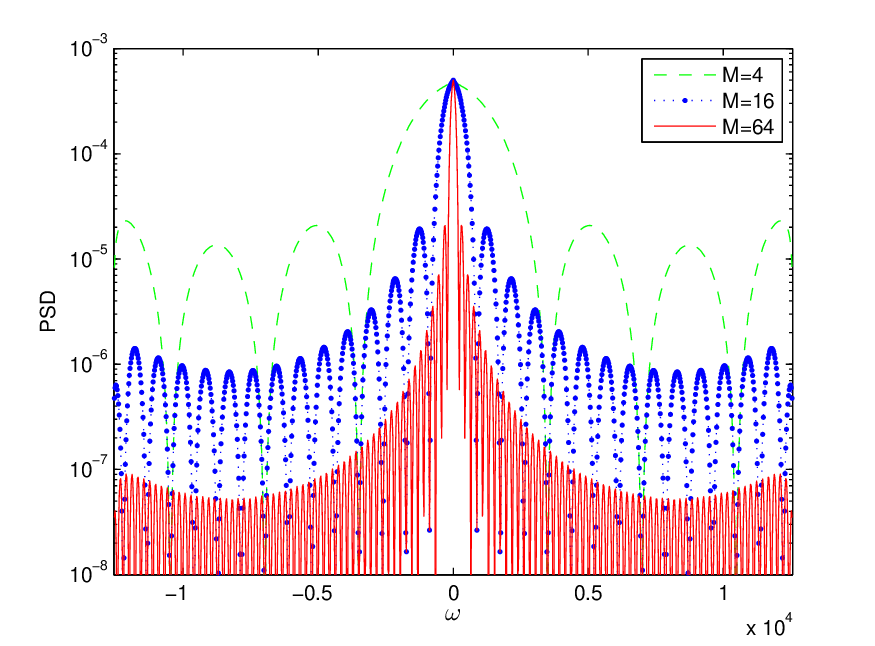}
\end{center}
\caption{ Comparison of channel PSD $P(\omega)$ with ULA composed of $M=4,16,64$ antennas, under $d=0.45\lambda$ and ``Equi-cos'' beamforming directions. }
\end{figure}

First, we evaluate the impact of the number of antennas on the PSD in Fig. 7. The ULA with $M\!=\!4,16,64$ antennas are considered. ``Equi-cos'' beamforming directions are adopted for all cases such that the beam-distortion function remains the same. Therefore, the exclusive contributing factor to the difference of the PSDs is the pattern function $|\mathcal{G}(\omega)|^{2}$, which in fact corresponds to the converted radiation pattern obtained with the MF beamformer pointing to the normal direction of ULA, as mentioned earlier. When the number of antennas $M$ increases, the radiation pattern exhibits lower sidelobe levels and narrower main and side lobes. These features are all reflected by the pattern function and thus by the PSDs depicted in Fig. 7. Since the sidelobes of the PSD cover the undesired high Doppler frequencies, a larger number of antennas can better reduce the sidelobe levels and thereby lead to smaller Doppler spread.

\begin{figure}[t]
\setlength{\abovecaptionskip}{-1mm}
\setlength{\belowcaptionskip}{-5mm}
\begin{center}
\includegraphics[width=80mm]{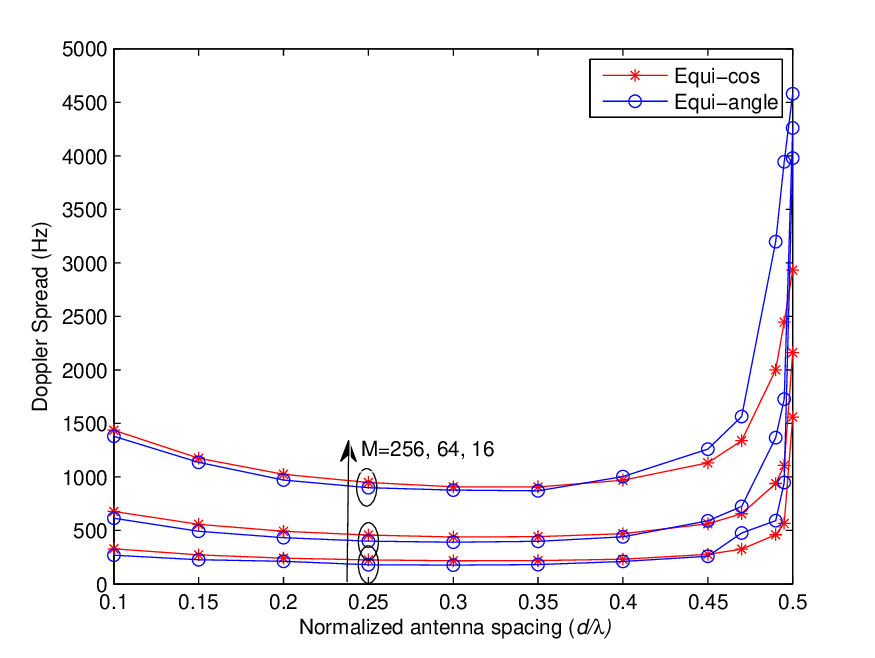}
\end{center}
\caption{ Comparison of Doppler spread ${{\sigma }_{\mathrm{DS}}}$ calculated by (\ref{DopplerSpread}), when the beamforming directions are configured in two different ways, with $M=16,64,256$ and $\frac{d}{\lambda} = [0.1, 0.15, 0.2, 0.25, 0.3, 0.35, 0.4, 0.45, 0.47, 0.49, 0.495, 0.5]$. }
\end{figure}

Then, the Doppler spreads computed by (\ref{DopplerSpread}) are compared in Fig. 8 under a set of normalized antenna spacings $\frac{d}{\lambda} \!=\! [0.1, 0.15, 0.2, 0.25, 0.3, 0.35, 0.4, 0.45, 0.47, 0.49, 0.495, 0.5]$, for different numbers of antennas $M\!=\!16,64,256$ and different configurations of beamforming directions ``Equi-cos'' (even $\cos \vartheta_q$) and ``Equi-angle'' (even $\vartheta_q$). The maximum Doppler shift is set as $f_d \!=\! 5\!\ 000\ \mathrm{Hz}$. The following observations can be drawn from Fig. 8:

1) The Doppler spread decreases with the increasing number of antennas $M$, which is within expectation. Actually, an enlarged antenna array provides higher spatial resolution and thereby the residual Doppler shifts tend to vanish, and the time variation of the equivalent channel after Doppler shifts compensation and transmit beamforming can be significantly alleviated.

2) There exists an optimal antenna spacing $d_{\mathrm{opt}}$ which yields the minimal Doppler spread, and an antenna spacing $d$ either smaller or larger than $d_{\mathrm{opt}}$ would be detrimental to appeasing the residual channel time variation. A too small $d$ cannot fully exploit the spatial resolution of the ULA, which is unfavorable for reducing the residual Doppler shifts. As $d$ increases to $0.5$, the aliasing between $0^{\circ} $ and $180^{\circ}$ would considerably enhance the PSD at high Doppler frequencies. Both factors contribute to large Doppler spread.

3) The Doppler spread is not sensitive to how the beamforming directions are configured for $\frac{d}{\lambda} \!\le\! 0.45$. However, different configurations of beamforming directions ``Equi-cos'' and ``Equi-angle'' have great impact on the Doppler spread for $\frac{d}{\lambda} \!=\! 0.5$. The significant divergence between the Doppler spread of ``Equi-cos'' and that of ``Equi-angle'' for $\frac{d}{\lambda} \!=\! 0.5$ can be explained as follows.
As previously mentioned, the pattern function $| \mathcal{G}(\omega)|^{2}$ accomplishes a full period within $(-2\omega_d, 2\omega_d)$ under $\frac{d}{\lambda} \!=\! 0.5$, which implies $\mathcal{G}(\pm 2\omega_d) \!=\! \mathcal{G}(0)$. Considering that the beam-distortion function of ``Equi-cos'' approaches 0 while that of ``Equi-angle'' approaches 1 as $\omega$ gets closer to $\pm 2\omega_d$, the PSD of ``Equi-cos'' at undesired high Doppler frequencies would be much smaller than ``Equi-angle''. Thus, the former attenuates the time variation of the equivalent channel more, resulting in smaller Doppler spread.

\subsection{Superiority of the Proposed Beamforming Network Optimization Technique}

In this subsection, we demonstrate numerically the superiority of the proposed CCAP beamforming network optimization technique, in terms of Doppler spread and uncoded symbol error rate (SER).

\begin{figure}[t]
\setlength{\abovecaptionskip}{-1mm}
\setlength{\belowcaptionskip}{-5mm}
\begin{center}
\includegraphics[width=80mm]{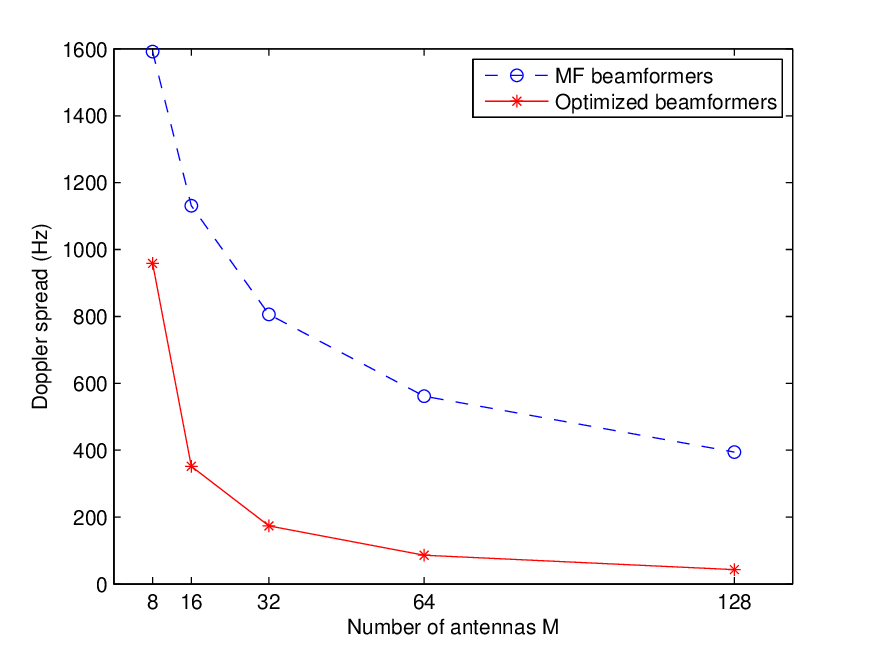}
\end{center}
\caption{ Comparison of the Doppler spreads computed with MF beamformers and optimized CCAP beamformers (i.e., (\ref{DopplerSpread}) and (\ref{DopplerSpread_CCAP})), under $M=8,16,32,64,128$, $\frac{d}{\lambda} = 0.45$ and ``Equi-cos'' beamforming directions. }
\end{figure}

In Fig. 9, we assess the effect of the proposed beamforming network optimization technique on reducing the Doppler spread, under ULA with different numbers of antennas $M\!=\!8, 16, 32, 64, 128$. The maximum Doppler shift is set as $f_d \!=\! 5\!\ 000\ \mathrm{Hz}$, and the beamforming directions satisfy ``Equi-cos'' configuration. The acquired optimal CCAP parameter $\hat{\mathbf{u}}$ for different numbers of antennas $M \!=\! 8,16,32,64 $ are shown in Table~\ref{CCAP}. Note that the maximum absolute value of $\hat{\mathbf{u}}$ is normalized to 1.
It can be seen from Fig. 9 that in contrast to the case with MF beamformers, the optimized beamformers ${\hat{\mathbf{b}}_{\mathrm{CCAP}}}\left( {{\vartheta }_{q}} \right)$ incorporating the CCAP parameter $\hat{\mathbf{u}}$ can substantially reduce the Doppler spread and thus suppress the residual channel time variation, regardless of the number of antennas. The better reduction of Doppler spread originates from the more effective attenuation of high Doppler frequencies, since the pattern function $|{\mathcal{G}}_{\mathrm{CCAP}}(\omega)|^{2}$ obtained with the optimized beamformers has much lower average SMR, as shown in Fig. 5.

After Doppler shifts compensation and multi-branch transmit beamforming (whether with MF beamformers or optimized beamformers ${\hat{\mathbf{b}}_{\mathrm{CCAP}}}\left( {{\vartheta }_{q}} \right)$), the received signal only suffers from slight residual time variation. Thus, the conventional channel estimation and data detection for time-invariant channel can be directly performed. Fig. \!10 compares the SER performance obtained with the received signals after transmit beamforming with MF beamformers and optimized beamformers, respectively. The receiver employs a 4-element ULA with antenna spacing $\frac{d}{\lambda} \!=\! 0.5$ and maximum-ratio-combining (MRC) receiver is used to detect the data symbols.
The transmitter is equipped with a large-scale ULA with $\frac{d}{\lambda} \!=\! 0.45$, and $M\!=\!32, 64, 128$ transmit antennas are considered. Note that despite of the number of transmit antennas $M$, the total average transmit power is always normalized to 1.
Each OFDM frame consists of 5 blocks, with the first block serving as pilot block. The number of subcarriers is taken as $N\!=\!128$, and both pilot and data symbols are randomly drawn from 16-QAM constellation. The maximum Doppler shift is set as $f_d \!=\! 1\!\ 000\ \mathrm{Hz}$ and the block duration is assumed to be $T_b\!=\!0.1 \mathrm{ms}$, which implies that the normalized maximum Doppler shift is $f_d T_b \!=\!0.1$. Moreover, the beamforming directions satisfy ``Equi-cos'' configuration.

Conventional BEM methods which directly tackle the time-varying channel without Doppler shifts compensation, including complex exponential BEM (CE-BEM) [7] and polynomial BEM (P-BEM) [8], are also included for comparison.
Single transmit antenna is considered for BEM methods, while the total average transmit power is kept to 1 to ensure the fairness of comparison.
Besides, an additional pilot block is appended at the end of each frame for BEM methods, to better capture the channel time variation~\cite{Rabbi_IETC2010}. The order of basis functions are taken as $N_{\mathrm{order}} =$ $4$.
A benchmark method, labelled as `NoDoppler-ML' is also provided. The maximum Doppler shift is 0, and similar to BEM methods, only a single transmit antenna is configured. Maximum likelihood (ML) channel estimator and MRC detector are employed to estimate the channel and detect the transmitted data, respectively.

\begin{table*}[!t]
\caption{ Optimal CCAP parameter obtained for different numbers of antennas $M=8,16,32,64$ }\label{CCAP}
\vspace{1.0em}
\centering
\begin{tabular}{|c||c|}
\hline
$M $ & Normalized CCAP parameter $\hat{\mathbf{u}}$ \\
\hline
$8 $ & 0.384, 0.656, 0.876, 1.000, 1.000, 0.876, 0.656, 0.384  \\
\hline
$16 $ & 0.106, 0.221, 0.364, 0.525, 0.687, 0.832, 0.941, 1.000, 1.000, 0.941, 0.832, 0.687, 0.525, 0.364, 0.221, 0.106 \\
\hline
$\multirow{2}*{32} $ & 0.060, 0.125, 0.207, 0.300, 0.399, 0.497, 0.591, 0.675, 0.748, 0.810, 0.863, 0.907, 0.943, 0.971, 0.990, 1.000, \vspace{-0.4em} \\
$~ $ & 1.000, 0.990, 0.971, 0.943, 0.907, 0.863, 0.810, 0.748, 0.675, 0.591, 0.497, 0.399, 0.300, 0.207, 0.125, 0.060 \\
\hline
$\multirow{4}*{64} $ & 0.030, 0.063, 0.104, 0.153, 0.206, 0.261, 0.314, 0.364, 0.410, 0.454, 0.494, 0.534, 0.573, 0.613, 0.652, 0.691, \vspace{-0.4em} \\
$~ $ & 0.727, 0.761, 0.792, 0.821, 0.847, 0.871, 0.893, 0.914, 0.934, 0.952, 0.967, 0.979, 0.988, 0.994, 0.998, 1.000, \vspace{-0.4em} \\
$~ $ & 1.000, 0.998, 0.994, 0.988, 0.979, 0.967, 0.952, 0.934, 0.914, 0.893, 0.871, 0.847, 0.821, 0.792, 0.761, 0.727, \vspace{-0.4em} \\
$~ $ & 0.691, 0.652, 0.613, 0.573, 0.534, 0.494, 0.454, 0.410, 0.364, 0.314, 0.261, 0.206, 0.153, 0.104, 0.063, 0.030 \\
\hline
\end{tabular}
\vspace{-1.0em}
\end{table*}

\begin{figure}[t]
\setlength{\abovecaptionskip}{-1mm}
\setlength{\belowcaptionskip}{-5mm}
\begin{center}
\includegraphics[width=80mm]{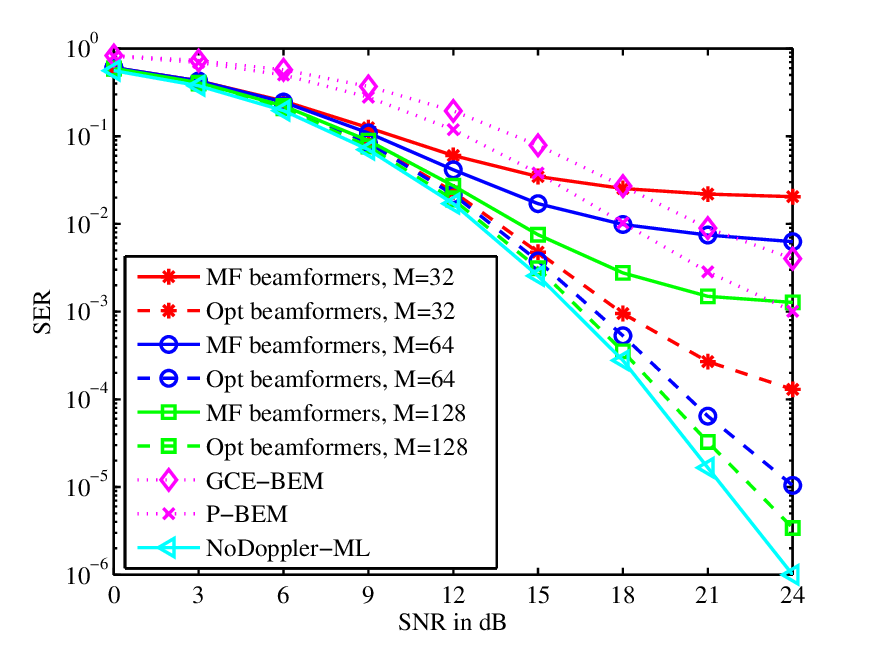}
\end{center}
\caption{ SER performance comparison between our beamforming-based approaches (`MF beamformers' and `Opt beamformers'), BEM methods (`GCE-BEM' and `P-BEM') and benchmark method `NoDopplerML', with 4-element receive ULA and normalized total transmit power. }
\end{figure}

From Fig. 10, the superiority of the proposed beamforming network optimization technique over the scheme using MF beamformers~\cite{Guo_GLOBECOM2017} is evident. Even with $M\!=\!128$ transmit antennas, the scheme with MF beamformers suffers from severe SER performance floor, which can be attributed to the residual channel time variation caused by uncompensated Doppler shifts. In fact, numerical results in~\cite{Guo_GLOBECOM2017} reveal that only when the number of transmit antennas are increased to $M\!=\!1024$, would the residual channel time variation become negligible and the SER performance floor disappear.
In contrast, the SER obtained with the optimized beamformers  ${\hat{\mathbf{b}}_{\mathrm{CCAP}}}\left( {{\vartheta }_{q}} \right)$ does not exhibit obvious floor even with $M\!=\!64$ and $128$ transmit antennas. This is due to the fact that compared to MF beamformers, the optimized CCAP beamformers can achieve more substantial reduction of Doppler spread, significantly improving the SER performance. In other words, with the proposed beamforming network optimization technique, far fewer transmit antennas are required to attain the same detection performance as MF beamformers.

In addition, the following observations could also be made from Fig. 10:
1) BEM methods can overmatch the scheme with MF beamformers in high SNR regions.
2) By introducing the CCAP parameter and optimizing the beamforming network, our beamforming-based approach (i.e., `Opt beamformers') remarkably outperforms BEM methods.
3) With the increasing number of transmit antennas, the proposed scheme `Opt beamformers' can gradually approach the benchmark method `NoDoppler-ML', confirming the effectiveness of the beamforming optimization technique in suppressing the residual Doppler shifts.

\section{Conclusions}
In this paper, we considered the angle-domain Doppler shifts compensation scheme for high-mobility uplink communication and derived the exact PSD and Doppler spread as a measure of assessing the residual channel time variation. The analysis reveals that the channel PSD can be fully characterized by the pattern function and beam-distortion function, which depend on the antenna spacing and the beamforming directions, respectively. Based on the delicately derived PSD with an explicit expression, the impacts of some essential parameters including antenna spacing and beamforming directions on channel PSD were discussed.
Moreover, in order to gain more effective reduction of Doppler spread, we further introduced the CCAP parameter to optimize the original MF beamforming network. The optimized beamformers incorporating CCAP parameter can ulteriorly suppress the residual Doppler shifts and thus yield slighter channel time variation.
Numerical results were provided to corroborate the channel PSD analysis and the proposed CCAP beamforming network optimization technique.

\appendices
\section{Derivation of the alternative form of the beam-distortion function (\ref{DistortionFunction_EquiAngle})}{\label{DerivationBDF}}
For $0 \!<\! |\tilde{\omega}| \!<\! 2$, the beam-distortion function (\ref{DistortionFunction_EquiAngle}) can be equivalently transformed into the following elliptic integral
\begin{align} {\label{EllipticIntegral}}
\mathcal{W}\left( {\tilde{\omega }} \right) = & \ \! \frac{2}{\mathrm{\pi}} \int_{0}^{\arccos \left( |\tilde{\omega}|-1 \right)}{\frac{1}{\sqrt{1-{{\left( \cos \vartheta -|\tilde{\omega}| \right)}^{2}}}}d\vartheta }, \nonumber \\
& \hspace{-1.6em} \overset{x=\cos \vartheta }{\mathop{=}}\, \frac{2}{\mathrm{\pi}} \int_{|\tilde{\omega }|-1}^{1}{\frac{1}{\sqrt{1-{{x}^{2}}}\sqrt{1-{{\left( x-|\tilde{\omega}| \right)}^{2}}}}dx}, \nonumber \\
& \hspace{-0.7em} \overset{*}{\mathop{=}}\,\ \! \frac{2}{\mathrm{\pi}} F\left( \frac{{\mathrm{\pi}}}{2}, \upsilon \right)= \frac{2}{\mathrm{\pi}} \int_{0}^{\frac{{\mathrm{\pi}}}{2}}{\frac{1}{\sqrt{1-{{\upsilon }^{2}}{{\sin }^{2}}\xi }}d\xi },
\end{align}
where $\overset{*}{\mathop{=}}\, \!$ employs the property of equation (3.147-4) in~\cite{Gradshteyn2007}, $\upsilon \!=\! \sqrt{1 \!-\! \frac{{{{\tilde{\omega }}}^{2}}}{4}}$ and $F\left( \psi ,k \right)$ is the elliptic integral of the first kind defined as
\begin{align*}
F\left( \psi ,k \right)=\int_{0}^{\psi }{\frac{1}{\sqrt{\left( 1-{{k}^{2}}{{\sin }^{2}}\alpha  \right)}}d\alpha }.
\end{align*}

Based on (\ref{EllipticIntegral}), we obtain that as $|\tilde{\omega}|$ goes to 2, $\upsilon $ approaches to 0. Thus, there holds
\begin{align} {\label{DistortionConvergeTo1}}
\underset{\left| {\tilde{\omega }} \right|\to 2}{\mathop{\lim }}\,\mathcal{W}\left( {\tilde{\omega }} \right) & =\frac{2}{\mathrm{\pi} } \int_{0}^{\frac{\mathrm{\pi} }{2}} {\underset{\upsilon \to 0} {\mathop{\lim }}\, \frac{1} {\sqrt{1-{{\upsilon }^{2}}{{\sin }^{2}}\xi }}d\xi }, \nonumber \\
& = \frac{2}{\mathrm{\pi} }\int_{0}^{\frac{\pi }{2}}{1d\xi }=1.
\end{align}

As for $\tilde{\omega} \!=\! 0$, there holds
\begin{align} {\label{DistortionIsInf}}
 \mathcal{W}\! \left( 0 \right)& = \! \frac{2}{\mathrm{\pi}}\! \int_{0}^{\mathrm{\pi}} \! {\frac{1}{\sqrt{1-{{\cos }^{2}} \vartheta }} d\vartheta }= \frac{2}{\mathrm{\pi}} \! \int_{0}^{\mathrm{\pi}} \! {\frac{1}{\sin \vartheta }d\vartheta }, \nonumber \\
 & = \frac{4}{\mathrm{\pi}} \! \int_{0}^{\frac{{\mathrm{\pi}}}{2}} \! {\frac{1}{\sin \vartheta }d\vartheta }= \frac{4}{{\mathrm{\pi}}} \! \int_{0}^{\frac{\mathrm{\pi}}{2}} \! {\frac{1}{\sin \frac{\vartheta }{2}\cos \frac{\vartheta }{2}} d\frac{\vartheta }{2}}, \nonumber \\
 & \! \overset{\tilde{\vartheta }=\frac{\vartheta }{2}}{\mathop{=}}\,\frac{4}{{\mathrm{\pi}}} \! \int_{0}^{\frac{{\mathrm{\pi}}}{4}} \! {\frac{1}{\tan \tilde{\vartheta }{{\cos }^{2}} \tilde{\vartheta }}d\tilde{\vartheta }} = \frac{4}{{\mathrm{\pi}}} \! \int_{0}^{\frac{{\mathrm{\pi}}}{4}} \! {\frac{1}{\tan \tilde{\vartheta }}d\tan \tilde{\vartheta }}, \nonumber \\
 & \hspace{-2mm} \overset{\gamma =\tan \tilde{\vartheta }}{\mathop{=}}\,\ \! \frac{4}{{\mathrm{\pi}}} \! \int_{0}^{1} \! {\frac{1}{\gamma }d\gamma }=\frac{4}{{\mathrm{\pi}}} \! \left. \ln \gamma  \right|_{0}^{1}=+\infty.
\end{align}

As a result, the beam-distortion function (\ref{DistortionFunction_EquiAngle}) can be alternatively expressed as (\ref{DistortionFunction_EquiAngle3}).


\begin{thebibliography}{99}

\bibitem{Wu_Access2016}
J. Wu and P. Fan, ``A survey on high mobility wireless communications: Challenges, opportunities and solutions,'' \emph{IEEE Access}, vol. 4, pp. 450-476, 2016.

\bibitem{He_VTM2016}
R. He, B. Ai, G. Wang, K. Guan, Z. Zhong, A. F. Molisch, C. Briso-Rodriguez, and C. P. Oestges, ``High-speed railway communications: From GSM-R to LTE-R,'' \emph{IEEE Veh. Technol. Mag.}, vol. 11, no. 3, pp. 49-58, Sep. 2016.

\bibitem{Zhou_TWC2015}
W. Zhou, J. Wu, and P. Fan, ``High mobility wireless communications with Doppler diversity: Fundamental performance limits,'' \emph{IEEE Trans. Wireless Commun.}, vol. 14, no. 12, pp. 6981-6992, Dec. 2015.

\bibitem{Hwang_TVT2009}
T. Hwang, C. Yang, G. Wu, S. Li, and G. Y. Li, ``OFDM and its wireless applications: A survey,'' \emph{IEEE Trans. Veh. Technol.}, vol. 58, no. 4, pp. 1673-1694, May 2009.

\bibitem{Wang_Access2018}
X. Wang, G. Wang, R. Fan, and B. Ai, ``Channel estimation with expectation maximization and historical information based basis expansion model for wireless communication systems on high speed railways,'' \emph{IEEE Access}, vol. 6, pp. 72-80, 2018.

\bibitem{Wang_TWC2011}
G. Wang, F. Gao, W. Chen, and C. Tellambura, ``Channel Estimation and Training Design for Two-Way Relay Networks in Time-Selective Fading Environments,'' \emph{IEEE Trans. Wireless Commun.}, vol. 10, no. 8, pp. 2681-2691, Aug. 2011.

\bibitem{Qu_TWC2010}
F. Qu and L. Yang, ``On the estimation of doubly-selective fading channels,'' \emph{IEEE Trans. Wireless Commun.}, vol. 9, no. 4, pp. 1261-1265, Apr. 2010.

\bibitem{Hijazi_TVT2009}
H. Hijazi and L. Ros, ``Polynomial estimation of time-varying multipath gains with intercarrier interference mitigation in OFDM systems,'' \emph{IEEE Trans. Veh. Technol.}, vol. 58, no. 1, pp. 140-151, Jan. 2009.

\bibitem{Souden_TSP2009}
M. Souden, S. Affes, J. Benesty, and R. Bahroun, ``Robust Doppler spread estimation in the presence of a residual carrier frequency offset,'' \emph{IEEE Trans. Signal Process.}, vol. 57, no. 10, pp. 4148-4153, Oct. 2009.

\bibitem{Tsai_SPL2009}
Y. R. Tsai and K. J. Yang, ``Approximate ML Doppler spread estimation over flat Rayleigh fading channels,'' \emph{IEEE Signal Process. Lett.}, vol. 16, no. 11, pp. 1007-1010, Nov. 2009.

\bibitem{Norklit_TVT1999}
O. Norklit and R. G. Vaughan, ``Angular partitioning to yield equal Doppler contributions,'' \emph{IEEE Trans. Veh. Technol.}, vol. 48, no. 5, pp. 1437-1442, Sep. 1999.

\bibitem{Chizhik_TWC2004}
D. Chizhik, ``Slowing the time-fluctuating MIMO channel by beam forming,'' \emph{IEEE Trans. Wireless Commun.}, vol. 3, no. 5, pp. 1554-1565, Sep. 2004.

\bibitem{Zhang_ICST2011}
Y. Zhang, Q. Yin, P. Mu, and L. Bai, ``Multiple Doppler shifts compensation and ICI elimination by beamforming in high-mobility OFDM systems,'' in \emph{Proc. Int. ICST Conf. Commun. and Netw. in China}, Aug. 2011, pp. 170-175.

\bibitem{Guo_ICSPCC2013}
W. Guo, P. Mu, Q. Yin, and H. M. Wang, ``Multiple Doppler frequency offsets compensation technique for high-mobility OFDM uplink,'' in \emph{Proc. IEEE ICSPCC}, Aug. 2013, pp. 1-5.

\bibitem{Rusek_SPM2013}
F. Rusek, D. Persson, B. K. Lau, E. G. Larsson, T. L. Marzetta, O. Edfors, and F. Tufvesson, ``Scaling up MIMO: opportunities and challenges with very large arrays,'' \emph{IEEE Signal Process. Mag.,} vol. 30, no. 1, pp. 40-60, Jan. 2013.

\bibitem{Larsson_CM2014}
E. Larsson, O. Edfors, F. Tufvesson, and T. Marzetta, ``Massive MIMO for next generation wireless systems,'' \emph{IEEE Commun. Mag.,} vol. 52, no. 2, pp. 186-195, Feb. 2014.

\bibitem{Ai_JSAC2017}
B. Ai, K. Guan, R. He, J. Li, G. Li, D. He, Z. Zhong, and K. M. S. Huq, ``On Indoor Millimeter Wave Massive MIMO Channels: Measurement and Simulation,'' \emph{IEEE J. Sel. Areas Commun.}, vol. 35, no. 7, pp. 1678-1690, Jul. 2017.

\bibitem{Zhang_TWC2018}
W. Zhang, F. Gao, S. Jin, and H. Lin, ``Frequency synchronization for uplink massive MIMO systems,'' \emph{IEEE Trans. Wireless Commun.}, vol. 17, no. 1, pp. 235-249, Jan. 2018.

\bibitem{Wang_TSP2018}
B. Wang, F. Gao, S. Jin, H. Lin, and G. Y. Li, ``Spatial- and frequency-wideband effects in millimeter-wave massive MIMO systems,'' \emph{IEEE Trans. Signal Process.}, vol. 66, no. 13, pp. 3393--3406, Jul. 2018.

\bibitem{Guo_TVT2017}
W. Guo, W. Zhang, P. Mu, and F. Gao, ``High-mobility OFDM downlink transmission with large-scale antenna array,'' \emph{IEEE Trans. Veh. Technol.}, vol. 66, no. 9, pp. 8600-8604, Sep. 2017.

\bibitem{Ge_VTC2017}
Y. Ge, W. Zhang, and F. Gao, ``High-mobility OFDM downlink transmission with partly calibrated subarray-based massive uniform linear array,'' in \emph{Proc. IEEE VTC-Spring}, Jun. 2017, pp. 1-6.

\bibitem{Ge_TWC2019}
Y. Ge, W. Zhang, F. Gao, and H. Minn, ``Angle-Domain Approach for Parameter Estimation in High-Mobility OFDM With Fully/Partly Calibrated Massive ULA,'' \emph{IEEE Trans. Wireless Commun.}, vol. 18, no. 1, pp. 591-607, Jan. 2019.

\bibitem{Guo_GLOBECOM2017}
W. Guo, W. Zhang, P. Mu, F. Gao, and B. Yao, ``Angle-domain Doppler pre-compensation for high-mobility OFDM uplink with massive ULA,'' in \emph{Proc. IEEE GLOBECOM}, Dec. 2017, pp. 1-6.

\bibitem{Guo_TWC2019}
W. Guo, W. Zhang, P. Mu, F. Gao, and H. Lin, ``High-Mobility Wideband Massive MIMO Communications: Doppler Compensation, Analysis and Scaling Law,'' \emph{IEEE Trans. Wireless Commun.}, vol. 18, no. 6, pp. 3177-3191, Jun. 2019.

\bibitem{Jakes_Wiley1994}
W. C. Jakes and D. C. Cox, \emph{Microwave mobile communications}. Wiley-IEEE Press, 1994.

\bibitem{Zheng_TC2003}
Y. R. Zheng and C. Xiao, ``Simulation models with correct statistical properties for Rayleigh fading channels,'' \emph{IEEE Trans. Commun.}, vol. 51, no. 6, pp. 920-928, Jun. 2003.

\bibitem{Alkhateeb_JSTSP2014}
A. Alkhateeb£¬ and O. E. Ayach, and G. Leus, and R. W. Heath, ``Channel Estimation and Hybrid Precoding for Millimeter Wave Cellular Systems,'' \emph{IEEE J. Sel. Topics Signal Process.}, vol. 8, no. 5, pp. 831-846, Oct. 2014.

\bibitem{Wang_TIT2015}
G. Wang, Q. Liu, R. He, F. Gao, and C. Tellambura, ``Acquisition of Channel State Information in Heterogeneous Cloud Radio Access Networks,'' \emph{IEEE Wireless Commun.}, vol. 22, no. 3, pp. 100-107, Jun. 2015.

\bibitem{Bellili_TC2017}
F. Bellili, Y. Selmi, S. Affes, and A. Ghrayeb, ``A low-cost and robust maximum likelihood joint estimator for the Doppler spread and CFO parameters over flat-fading Rayleigh channels,'' \emph{IEEE Trans. Commun.}, vol. 65, no. 8, pp. 3467-3478, Aug. 2017.

\bibitem{You_TWC2015}
L. You, X. Gao, X. Xia, N. Ma, and Y. Peng, ``Pilot reuse for massive MIMO transmission over spatially correleated Rayleigh fading channels,'' \emph{IEEE Trans. Wireless Commun.}, vol. 14, no. 6, pp. 3352-3366, Jun. 2015.

\bibitem{Zhang2017}
X.-D. Zhang, \emph{Matrix Analysis and Applications, Cambridge}, U.K., Cambridge Univ. Press, Oct. 2017.

\bibitem{Lebret_TSP1997}
H. Lebret and S. Boyd, ``Antenna array pattern synthesis via convex optimization,'' \emph{IEEE Trans. Signal Process.,} vol. 45, no. 3, pp. 526-532, Mar. 1997.

\bibitem{Shi_SPL2005}
Z. Shi and Z. Feng, ``A new array pattern synthesis algorithm using the two-step least-squares method,'' \emph{IEEE Signal Process. Lett.,} vol. 12, no. 3, pp. 250-253, Mar. 2005.

\bibitem{Wang_TWC2016}
P. Wang, Y. Li, Y. Peng, S. C. Liew, and B. Vucetic, ``Non-Uniform Linear Antenna Array Design and Optimization for Millimeter-Wave Communications,'' \emph{IEEE Trans. Wireless Commun.,} vol. 15, no. 11, pp. 7343-7356, Nov. 2016.

\bibitem{Rabbi_IETC2010}
M. F. Rabbi, S. W. Hou, and C. C. Ko, ``High mobility orthogonal frequency division multiple access channel estimation using basis expansion model,'' \emph{IET Commun.}, vol. 4, no. 3, pp. 353-367, Feb. 2010.

\bibitem{Gradshteyn2007}
I. S. Gradshteyn and I. M. Ryzhik, \emph{Table of integrals, series, and products}. New York, NY, USA: Academic, 2007.

\end{thebibliography}
\end{document}